\documentstyle[12pt,bezier,epsfig]{article}
\begin{document}
\def \inbar{\vrule height1.5ex width.4pt depth0pt}
\def \xC{\relax\hbox{\kern.25em$\inbar\kern-.3em{\rm C}$}}
\def \xR{\relax{\rm I\kern-.18em R}}
\newcommand{\xZ}{Z \hspace{-.08in}Z}
\newcommand{\xbe}{\begin{equation}}
\newcommand{\xee}{\end{equation}}
\newcommand{\xbea}{\begin{eqnarray}}
\newcommand{\xeea}{\end{eqnarray}}
\newcommand{\xnn}{\nonumber}
\newcommand{\xkt}{\rangle}
\newcommand{\xbr}{\langle}
\newcommand{\xlll}{\left( }
\newcommand{\xrrr}{\right)}
\newcommand{\xcun}{\mbox{\footnotesize${\cal N}$}}

\title{Adiabatic Geometrical Phase for Scalar  Fields 
in a Curved Spacetime}
\author{Ali Mostafazadeh\thanks{E-mail: 
alimos@phys.ualberta.ca}\\ \\
Theoretical Physics Institute, University of Alberta, \\
Edmonton, Alberta  T6G 2J1,  Canada, \\
and\\
Department of Mathematics, College of Arts and Sciences,\\
Ko\c{c} University,
Istinye, 80860 Istanbul, Turkey. }
\date{June 1997}
\maketitle

\begin{abstract}
A convenient  framework is developed to generalize Berry's
investigation of the adiabatic geometrical phase for a classical
relativistic charged scalar field in a curved background  spacetime
which is minimally coupled to electromagnetism and an arbitrary
(non-electromagnetic) scalar potential.  It involves a
two-component formulation of the corresponding Klein-Gordon
equation. A precise definition of the adiabatic approximation is
offered and conditions of its validity are discussed. It is shown 
that the adiabatic geometric phase can be computed without
making a particular choice for an inner product on the space of
solutions of the field equations. What is needed is just an inner
product on the Hilbert space of the square integrable functions
defined on the spatial hypersurfaces. The two-component formalism
is applied in the investigation of the adiabatic geometric phases
for several specific examples, namely, a rotating magnetic field in
Minkowski space, a rotating cosmic string, and an arbitrary
spatially homogeneous cosmological background. It is shown that
the two-component formalism reproduces the known results for the
first two examples. It also leads to several interesting results for
the case of  spatially homogeneous cosmological models. In
particular, it is shown that the adiabatic geometric phase angles
vanish for Bianchi type I models. The situation is completely
different for Bianchi type IX models where a variety of nontrivial
non-Abelian adiabatic geometrical phases can occur. The analogy
between the adiabatic geometric phases induced by the Bianchi
type IX backgrounds and those associated with the well-known
time-dependent nuclear quadrupole Hamiltonians is also pointed
out. 
\end{abstract}
Number of pages =  39 + one figure

\newpage

\baselineskip=24pt

.
\vspace{3cm}

\begin{itemize}
	\item
{\bf Proposed Running head:} Geometric phase of scalar fields
	\item
{\bf Author:} Ali Mostafazadeh
	\item
{\bf Contact Information:}\\
	\begin{itemize}
	\item
{\bf Before June 27, 1997:}\\
\underline{Address}: 412 Physics Lab, University of Alberta, 
Edmonton, AB, T6G 2J1, Canada\\
\underline{Phone}: (403) 434-5877, (403) 492-3960\\
\underline{Fax}: (403) 492-0714
	\item
{\bf After June 27, 1997:}\\
\underline{Address}: Department of Mathematics, College
of Arts and Sciences, Ko\c{c} University, \c{C}ayir Cad.~no:~5,
Istinye, 80860 Istanbul, Turkey\\
\underline{Fax}: (90-212) 229-0680
	\item
{\bf E-mail:}  alimos@phys.ualberta.ca
	\end{itemize}
	\end{itemize}

\newpage

\section{Introduction}
Following the pioneering works of Berry \cite{berry1984},
Wilczek and Zee \cite{wi-ze} and Aharonov and Anandan
\cite{aa} on geometric phases in non-relativistic quantum mechanics,
Garrison and Chiao \cite{ga-ch} showed that the geometric phases
may be defined for any classical field theory provided that some
gauge symmetries were present. The role of the gauge symmetry
was to provide a conserved charge which in turn defined an inner
product on the function space of the classical fields.
Anandan \cite{an}, commented on the latter article, indicating that
the condition of gauge symmetry may in general be relaxed or
practically replaced with the condition of the existence of an inner
product on the space of fields. A common assumption of both
Refs.~\cite{ga-ch} and \cite{an} was that the field equations involved
only first time derivatives of the fields which is really not a
restriction. 

Clearly the simplest  classical field theories of interest are the
Klein-Gordon fields on the ordinary Minkowski spacetime. The
phenomenon of the geometric phase for charged Klein-Gordon
fields minimally coupled to a time-dependent electromagnetic field
has already been studied by Anandan and Mazur \cite{an-ma}.
The main strategy of \cite{an-ma} is to decompose the vector space
of the fields into three subspaces which are spanned respectively by
the positive, zero,  and negative frequency (energy) solutions
and to note that on the positive and negative frequency subspaces,
where the Klein-Gordon inner product is positive, respectively,
negative definite, the  Klein-Gordon equation may be written as a
pair of equations which are linear in the time-derivative of the field.
More recently, a similar approach has been pursued to study the
dynamics of Klein-Gordon fields in a periodic
Friedmann-Robertson-Walker background by Droz-Vincent \cite{dr}.

The idea of investigating the manifestations of  the geometric phase
in the context of gravitation and cosmology is in fact not quite new.
The first developments in this direction, to best of my knowledge,
goes back to the work of Brout and Venturi \cite{br-ve} which
was inspired by the earlier results of Banks \cite{ba} and Brout
\cite{br} on the use of Born-Oppenheimer approximation in
semiclassical treatment of the Wheeler-DeWitt equation, and the
application of Berry's phase in improving the Born-Oppenheimer
approximation in molecular physics \cite{mo-sh-wi}. Subsequent work
which followed essentially the same idea is that of Venturi
\cite{ve1,ve2}, Casadio and Venturi  \cite{ca-ve} and Datta \cite{da}.
There is also the contributions of  Cai and Papini \cite{ca-pa}
which are based on the four-space-formalism of the relativistic
quantum mechanics.
More recently Corichi and Pierri \cite{co-pi} considered Klein-Gordon
fields in a class of stationary spacetimes and in particular 
investigated the induced topological Aharonov-Bohm type phases
due to a rotating cosmic string. The analogy between the
topological phase due to a rotating cosmic string and the
Aharonov-Bohm phase had previously been pointed out by
de~Sousa Gerbert and Jackiw \cite{ge-ja}.

The purpose of this article is to study the geometric phases
associated with a charged Klein-Gordon field $\Phi$ in an arbitrary
globally hyperbolic spacetime  $(M,{\rm g})$ which is minimally
coupled to an electromagnetic potential $A$, as well as an arbitrary
scalar potential $V$. The latter may, for instance, be identified with
the appropriate multiple of the Ricci scalar curvature which renders
the theory conformally invariant. The problem of investigation of the
dynamics of such a field theory has a long history in the context of
developing quantum field theories in a curved background
\cite{bd,fu,wa}. However, I shall not be concerned with subtleties
associated with the full second quantized theory. Instead, the
Klein-Gordon field will be viewed and treated as a classical (first
quantized) field.

In section 2, the two-component form of the field equation
is derived. This allows for a simple application of the adiabatic
theorem which following Berry's original approach
\cite{berry1984} yields the definition of  the adiabatically cyclic
states and the associated dynamical and geometric phases for
the theory. This is described in section~3 and applied to the
problems of a rotating magnetic field in Minkowski space in
section~4, a rotating cosmic string in section~5, and spatially
homogeneous (Bianchi) cosmological backgrounds in
section~6. The latter section also includes a detailed analysis
of Bianchi type I and IX cases. Particularly interesting
is the analogy between the case of a Bianchi type IX background
and the quadrupole Hamiltonians of the molecular and nuclear
physics. 

\section{Two-Component Formalism}

Consider  complex scalar fields $\Phi$ defined on a globally 
hyperbolic spacetime $(M,{\rm g})=(\xR\times\Sigma,{\rm g})$
satisfying
	\xbe
	\left[g^{\mu\nu}(\nabla_\mu+ieA_\mu)(\nabla_\nu+
	ieA_\nu)+V-\mu^2\right]\Phi=0\;,
	\label{k-g-1}
	\xee
where $g^{\mu\nu}$ are components of the inverse
of the metric ${\rm g}$, $\nabla_\mu$ is the covariant derivative
along $\partial/\partial x^\mu$ defined by the Levi Civita
connection,  $A_\mu$ are components of the electromagnetic
potential, $V$ is an arbitrary scalar potential, $e$ is the electric
charge, and $\mu$ is the mass. Throughout this article the
signature of the metric ${\rm g}$ is chosen to be $(-,+,+,+)$,
letters from the beginning and the middle of the Greek alphabet
are associated with an arbitrary local basis and a local
coordinate basis of the tangent spaces (bundle) of the
spacetime manifold, respectively. The letters from the beginning
and the middle of the Latin alphabet label the corresponding
spatial components. They take $1,~2$ and $3$. 

Denoting a time derivative by a dot, one can express
Eq.~(\ref{k-g-1}) in the form:
	\xbe
	\ddot\Phi+\hat D_1\dot\Phi+\hat D_2\Phi=0\;,
	\label{fi-eq}
	\xee
where
	\xbea
	\hat D_1&:=&\frac{2}{g^{00}}\left[ g^{0i}\partial_i+
	ieg^{0\mu}A_\mu-
	\frac{1}{2}\:g^{\mu\nu}\Gamma_{\mu\nu}^0\right]\;,
	\label{d1}\\
	\hat D_2&:=&\frac{2}{g^{00}}\left[ \frac{1}{2}\:g^{ij}
	\partial_i\partial_j+
	(ieg^{\mu i}A_\mu-\frac{1}{2}\:g^{\mu\nu}
	\Gamma_{\mu\nu}^i)\partial_i+
	\right.\xnn\\
	&&\left.
	\frac{1}{2}\,g^{\mu\nu}(ie\nabla_\mu A_\nu-
	e^2A_\mu A_\nu)+
	\frac{1}{2}\:(V-\mu^2)\right]\,.
	\label{d2}
	\xeea

A two-component  representation of the field equation
(\ref{fi-eq}) is
	\xbe
	i\dot\Psi^{(q)}=\hat H^{(q)}\Psi^{(q)}\;,
	\label{sch-eq-q}
	\xee
where
	\xbea
	\Psi^{(q)}&:=&\left(\begin{array}{c} u^{(q)}\\v^{(q)}
	\end{array}\right)\,,
	\label{psi-q}\\
	u^{(q)}&:=&\frac{1}{\sqrt{2}}\:(\Phi+q\dot\Phi)\;,~~~
	v^{(q)}\::=\:\frac{1}{\sqrt{2}}\:(\Phi-q\dot\Phi)\;,	\label{u-v-q}\\
	\hat H^{(q)}&:=&\frac{i}{2}\left(
	\begin{array}{cc}
	\frac{\dot q}{q}+\frac{1}{q}-\hat D_1-q\hat D_2&
	-\frac{\dot q}{q}-\frac{1}{q}+\hat D_1-q\hat D_2\\
	&\\
	-\frac{\dot q}{q}+\frac{1}{q}+\hat D_1+q\hat D_2&
	\frac{\dot q}{q}-\frac{1}{q}-\hat D_1+q\hat D_2
	\end{array}\right)\,,
	\label{h-q}
	\xeea
and $q$ is an arbitrary, possibly time-dependent, non-zero complex
parameter. The set  $\xC-\{0\}$ of  $q$'s defines a
group of transformations
	\xbe
	\Psi^{(q)}\to\Psi^{(q')}=:g(q',q)\Psi^{(q)}
	\label{gauge-trans}
	\xee
which is isomorphic to $GL(1,\xC)$. The group elements are given by
	\[g(q',q)=g(\gamma)=\left(\begin{array}{cc}
	\frac{1+\gamma}{2}&\frac{1-\gamma}{2}\\
	\frac{1-\gamma}{2}&\frac{1+\gamma}{2}
	\end{array}\right)\;,\]
where $\gamma :=q'/q$. Under the transformation (\ref{gauge-trans}),  
the Hamiltonian transforms according to
	\[\hat H^{(q)}\to\hat H^{(q')}=
	g(\gamma)\hat H^{(q)}g^{-1}(\gamma)+ i\dot g(\gamma) 	g^{-1}(\gamma)\;,\] 
and the Schr\"odinger equation~(\ref{sch-eq-q}) preserves its form. 
The underlying $GL(1,\xC)$ symmetry which characterizes  the 
arbitrariness of $q$
does not have any physical significance. It is, however, useful for
computational purposes. A concrete example is given in section~6.

The advantage of the two-component form of the field
equation is that it enables one to proceed in a manner
analogous with the well-known non-relativistic quantum
mechanical case. Indeed Eq.~(\ref{sch-eq-q}) with a fixed
choice of $q$ is a Schr\"odinger equation associated with an
explicitly time-dependent Hamiltonian $\hat H^{(q)}$. The
two-component fields $\Psi^{(q)}$ belong to the vector space
${\cal H}_t\oplus{\cal H}_t$ where ${\cal H}_t$ is the Hilbert
space completion (with respect to an appropriate inner
product) of compactly supported complex-valued functions
on the spatial hypersurface $\Sigma_t$ associated with a
specific ADM decomposition of the spacetime \cite{mtw}. 

Usually in the two-component approach to the Klein-Gordon
field theory in  Minkowski spacetime, one chooses an inner
product on ${\cal H}_t\oplus{\cal H}_t$ in such a way as to
make the corresponding Hamiltonian self-adjoint
\cite{fe-vi,holstein}. A Hermitian  inner product  $(~,~)$ on
${\cal H}_t\oplus{\cal H}_t$ may be  defined by a Hermitian
inner product $\xbr~|~\xkt$ on ${\cal H}_t$ and a  possibly
time-dependent complex Hermitian $2\times 2$ matrix
$h=(h_{rs})$:
	\xbe
	(\Psi_1,\Psi_2):=(\xbr u_1|,\xbr v_1|)
	\left(\begin{array}{cc}
	h_{11}&h_{12}\\
	h_{12}^*&h_{22} \end{array}\right)
	\left(\begin{array}{c}
	|u_2\xkt\\
	|v_2\xkt \end{array}\right)\;,
	\label{he-in-pr}
	\xee
where $u_r$ and $v_r$ are components of $\Psi_r$, and $h_{11}$
and $h_{22}$ are real. The usual choice for $h$, in the Minkowski
case, is $h_{11}=-h_{22}=1,~h_{12}=0$, \cite{fe-vi,holstein}. This
choice leads to
	\xbe
	(\Psi_1,\Psi_2)=\xbr u_1|u_2\xkt-\xbr v_1|v_2\xkt\;.
	\label{he-in-pr-sp}
	\xee
It is not difficult to check that in the general case this choice does
not guarantee the self-adjointness of the Hamiltonian unless some
severe conditions are imposed on $q$ and the operators $\hat D_1$
and $\hat  D_2$, namely, that $q$ must be imaginary, $\hat  D_2$
must be self-adjoint with respect to the inner product $\xbr~|~\xkt$
on ${\cal H}_t$, and $\hat D_1=\dot q/q$. The latter condition is
especially restrictive as $q$ can only depend on time and being a
free (non-dynamical) parameter, may be set to a constant in which
case $\hat D_1$ must vanish.  In general, these conditions are
not fulfilled. Nevertheless, the inner product (\ref{he-in-pr-sp})
has an appealing property which is described next. 

Consider the eigenvalue problem for $H^{(q)}$. Denoting the
eigenvalues and eigenvectors by $E_n^{(q)}$ and $\Psi_n^{(q)}$,
i.e.,
	\xbe
	H^{(q)}\Psi_n^{(q)}=E_n^{(q)}\Psi_n^{(q)}\;,
	\label{eg-va-eq-q}
	\xee
expressing $\Psi_n^{(q)}$ in the two-component form, and using
Eq.~(\ref{h-q}), one can easily show that up to an undetermined
scalar multiple, $\Psi_n^{(q)}$ has the following form:
	\xbe
	\Psi_n^{(q)}=\frac{1}{\sqrt{2}}\:\left(\begin{array}{c}
	1-iqE_n^{(q)}\\
	1+iqE_n^{(q)}
	\end{array}\right)\: \Phi_n^{(q)}\;,
	\label{eg-ve}
	\xee
where $\Phi_n^{(q)}\in {\cal H}_t$ satisfies:
	\xbe
	\left[ \hat D_2-iE_n^{(q)}(\hat D_1-\frac{\dot q}{q})
	-\left( E_n^{(q)}\right)^2\right] \Phi_n^{(q)}=0\;.
	\label{ge-eg-va-eq}
	\xee
This equation may be viewed as a `generalized' eigenvalue
equation\footnote{Note that this terminology has nothing to
do with the concept of generalized eigenvalues of spectral
analysis.} in ${\cal H}_t$. It defines both the vectors $\Phi_n^{(q)}$
and the complex numbers $E_n^{(q)}$. It reduces to the ordinary
eigenvalue equation for $\hat D_2$, if $\hat D_1=\dot q/q$.
Note that this is also one of the conditions for  self-adjointness
of the Hamiltonian, with the choice of (\ref{he-in-pr-sp}) for the
inner product. Furthermore, if this condition is satisfied, then
Eq.~(\ref{ge-eg-va-eq}) determines $E_n^{(q)}$ up to a sign, i.e.,
eigenvalues come in pairs of opposite sign.

If $q$ is chosen to be time-independent, then (\ref{ge-eg-va-eq})
does not carry any information about $q$ and therefore
$\Phi_n^{(q)}$ and $E_n^{(q)}$ are independent of the choice
of $q$. \footnote{This can also be seen by noting that under the
transformation $\Psi_n^{(q)}\to \tilde
\Psi_n^{(q)}= g(q',q)\Psi_n^{(q)}$, the eigenvectors
preserve their form (\ref{eg-ve}) and that $\tilde\Psi_n^{(q)}$
is an eigenvector of $\hat H^{(q')}$ with the same eigenvalue
$E_n^{(q)}$.} Hence, one can drop the labels $(q)$
on the right hand side of  Eq.~(\ref{eg-ve}). In this case, 
Eq.~(\ref{ge-eg-va-eq}) becomes:
	\xbe
	\left[ \hat D_2-iE_n\hat D_1-E_n^2\right] \Phi_n=0\;.
	\label{ge-eg-va-eq-2}
	\xee

Now let us use the inner product (\ref{he-in-pr-sp}) to compute
the inner product of two eigenvectors of the Hamiltonian. 
Performing the algebra, one finds
	\xbe
	(\Psi_{m}^{(q)},\Psi_{n}^{(q)})=i(q^*E_m^*-qE_n)
	\xbr \Phi_m|\Phi_n\xkt\;.
	\label{orthogonality}
	\xee
Therefore if $q$ is a positive imaginary number, i.e., $q=i|q|$,
then
	\xbe
	(\Psi_{m}^{(q)},\Psi_{n}^{(q)})=|q|(E_m^*+E_n)
	\xbr \Phi_m|\Phi_n\xkt\;.
	\label{orthogonality-1}
	\xee
Hence the eigenvectors $\Psi_{n}^{(q)}$ and $\Psi_{m}^{(q)}$
with $E_m=-E_n^*$ (if they exist) are orthogonal regardless of
the value of $\xbr \Phi_m|\Phi_n\xkt$.  Furthermore, one has 
$(\Psi_{n}^{(q)},\Psi_{n}^{(q)})=2|q|{\rm Re}(E_n)\xbr\Phi_n|
\Phi_n\xkt$, i.e., the norm of an energy eigenvector has
the same sign as the real part of the corresponding eigenvalue.
It vanishes for the zero\footnote{The existence of a zero energy
eigenvector $\Psi_0^{(q)}$ is equivalent to the existence of the
solution to the differential equation $\hat D_2 \Phi_0=0$.} and
imaginary energy eigenvalues. Note that here I am assuming
that the inner product $\xbr~|~\xkt$ on ${\cal H}_t$ is
non-negative. In fact ${\cal H}_t$ is to be identified with the
separable Hilbert space $L^2(\Sigma_t)$ of square-integrable
functions on $\Sigma_t$ where the integration is defined by 
the measure $[\det(^{(3)}{\rm g})]^{1/2}$  defined by the
Riemannian three-metric $^{(3)}{\rm g}$ induced by the
four-metric ${\rm g}$.

Another interesting property of the inner product
(\ref{he-in-pr-sp}) is the fact that for imaginary $q$ it
yields the familiar Klein-Gordon inner product,
$\xbr~,~\xkt_{\rm KG}$. This is easily seen by substituting
(\ref{u-v-q}) in (\ref{he-in-pr-sp}), which leads to:
	\xbe
	(\Psi_1,\Psi_2)=q\xbr\Phi_1|\dot\Phi_2\xkt
	+q^*\xbr\dot\Phi_1|\Phi_2\xkt=q\left[ \xbr\Phi_1|
	\dot\Phi_2\xkt-\xbr\dot\Phi_1|\Phi_2\xkt\right]=:
	q\xbr\Phi_1,\Phi_2\xkt_{\rm KG}\;.
	\label{K-G-in-pr}
	\xee

It is also useful to recall
that the space ${\cal H}_t\oplus{\cal H}_t$ is nothing but the
space of the possible initial conditions $[\Phi(t,x^i),\dot\Phi(t,x^i)]$
with initial time being $t$ and $(x^i)\in\Sigma_t$. In view of the
well-posedness of the dynamical equation \cite{wald-gr}, this
(vector) space is isomorphic to the space of solutions of the
field equation (\ref{k-g-1}). Hence a two-component decomposition
may be viewed as a splitting of the space of solutions of the
field equations. In view of the freedom of choice of the
parameter $q$, this splitting is clearly not unique. 

\section{Cyclic States and Adiabatic Geometric Phase}
By definition a cyclic state (an element of the projective Hilbert
space) of a quantum mechanical system, whose dynamics
is governed by the  Schr\"odinger equation
	\xbe
	i \dot\psi(t)=\hat H(t)\: \psi(t)\;,
	\label{sch-eq}
	\xee
is said to be cyclic with a period $\tau$, if it is an eigenstate of
the time-evolution operator $\hat U(\tau):={\cal T}
\exp[-i\int_0^\tau \hat H(t)]$. Here ${\cal T}$ is
the time-ordering operator. An associated initial state vector
$\psi(0)$ then satisfies:
	\xbe
	\psi(\tau)=\hat U(\tau)\psi(0)=e^{i\alpha(\tau)}\psi(0)\;,
	\label{cyclic}
	\xee
where $\alpha(\tau)\in\xC$. If the Hamiltonian is self-adjoint, then
$\alpha(\tau)\in\xR$ and consequently $\psi(\tau)$ and $\psi(0)$
differ by a phase. In general $\alpha(\tau)$ may be expressed as
the sum of a dynamical and a geometrical part \cite{aa}.
This decomposition uses the inner product structure of the Hilbert
space.

The situation is rather more transparent if  the time-dependence
of the Hamiltonian is adiabatic. In this case, one can follow Berry's
approach \cite{berry1984} of employing the adiabatic theorem of
quantum mechanics. According to the adiabatic theorem
\cite{messiyah}, if the initial state is an eigenstate of the initial
Hamiltonian $\hat H(0)$, then after a time period $t>0$ it evolves
into an eigenstate of the Hamiltonian $\hat H(t)$. More precisely
assume that the time-dependence of the Hamiltonian is realized
through its dependence on a set of parameters $R=(R^1,\cdots,
R^n)$ and a smooth curve $C:[0,\tau]\to {\cal M}$, where $R$ is
viewed as coordinates of a parameter space ${\cal M}$, i.e.,
$\hat H(t):=H[R(t)]$, and $R(t)=(R^1(t),\cdots,R^n(t)):=C(t)$.
Furthermore, let $\psi_n[R]$ denote eigenvectors of $\hat H[R]$
with eigenvalue $E_n[R]$:
	\xbe
	\hat H[R]\: \psi_n[R]=E_n[R] \psi_n[R]\;,
	\label{eg-va-eq}
	\xee
whose dependence on $R$ is assumed to be smooth, the degree
of the degeneracy of the eigenvalues is independent of $R$, and
for $R=R(t)$ there is no level crossings. Then the statement of the
adiabatic theorem may be summarized by the
following (approximate) equation:
	\xbe
	\psi(t):=\hat U(t)\psi_n(0)\approx e^{i\alpha_n(t)}\psi_n(t)\;,
	\label{ad-ap}
	\xee
where $\psi_n(t):=\psi_n[R(t)]$.
If $E_n(t):=E_n[R(t)]$ is $\xcun$-fold degenerate, then $\psi_n$
belongs to the $\xcun$-dimensional degeneracy subspace
${\cal H}_n$ and $\alpha_n$ is an $\xcun\times \xcun$
matrix-valued function of time. The approximation sign $\approx$
in (\ref{ad-ap}) is used to remind one of the fact that this relation
is only valid if the adiabatic approximation is justified. 

Assuming the validity of the adiabatic approximation ($\approx
\to=$) and substituting (\ref{ad-ap}) in the Schr\"odinger equation
(\ref{sch-eq}), one has \cite{wi-ze}:
	\xbea
	e^{i\alpha_n(t)}&=&\exp [{-i\int_0^tE_n(t')dt'}]\;{\cal P}\,
	\exp[{i\int_{C(0)}^{C(t)}{\cal A}_n}]\;,
	\label{e^ia}\\
	{\cal A}_n^{IJ}[R]&:=&\frac{i \xbr \psi^I_n[R],\frac{\partial}{
	\partial R^a}\:
	\psi^J_n[R]\xkt}{\xbr\psi^I_n[R],\psi^I_n[R]\xkt}\: dR^a
	\:=\: \frac{i \xbr \psi^I_n[R],d \psi^J_n[R]\xkt}{\xbr\psi^I_n[R],
	\psi^I_n[R]\xkt}\;,
	\label{connection}
	\xeea
where ${\cal P}$ is the path-ordering operator, $\psi_n^I[R]$
form a complete orthonormal basis of
the degeneracy subspace ${\cal H}_n$, and $\xbr~,~\xkt$ is the
inner product. If the Hamiltonian is periodic, i.e., $C$ is a closed
curve, then according to  (\ref{ad-ap}), $\psi_n[R(0)]=\psi_n[R(T)]$
is a cyclic state vector. In this case the first and the second
(path-ordered) exponential in (\ref{e^ia}), with $t=T$, are called
the dynamical and the geometrical parts of the total adiabatic
matrix-valued phase $\exp[i\alpha_n(T)]$, respectively,
\cite{wi-ze}. The qualification `geometrical' is best justified by
identifying the geometric part of the phase by the holonomy of
a principal spectral bundle over the parameter space ${\cal M}$
or alternatively the universal classifying bundle over the projective
Hilbert space, \cite{si,aa,p6}.

The situation for a non-self-adjoint Hamiltonian is rather
more complicated. The dynamical and the geometrical phase
can still be defined  in terms of the projective Hilbert space
\cite{sa-bh}. However, in general the eigenvectors of the
Hamiltonian are not orthogonal.\footnote{Note that the
eigenvectors within a single degeneracy subspace can always
be orthonormalized. However the eigenvectors corresponding to
distinct eigenvalues in general overlap.} This renders the
proof of the adiabatic theorem \cite{messiyah} invalid.  One
can still postulate (\ref{ad-ap}) as an ansatz which may or may
not be valid for specific evolutions. The condition of
the validity of this ansatz, which allows one to pursue the
same strategy in defining the adiabatic geometric phase, is
	\xbe
	\xbr \psi_m,\psi_n\xkt\:
	\xbr\psi_n,\dot\psi_n\xkt  =
	\xbr\psi_m,\dot\psi_n\xkt\:
	\xbr \psi_n,\psi_n\xkt\;,
	\label{condition}
	\xee
where $\psi_m$ and $\psi_n$ are any pair of  distinct
eigenvectors of the Hamiltonian. This condition is obtained
by substituting the ansatz (\ref{ad-ap}) in the Schr\"odinger
equation (\ref{sch-eq}) and taking the inner product of both sides
of the resulting equation with $\psi_m$.  Eq.~(\ref{condition}) is 
trivially satisfied for the case of a self-adjoint Hamiltonian. In
this case, the left hand side vanishes identically and $\xbr\psi_m,
\dot\psi_n\xkt$ for $m\neq n$, vanishes approximately by
virtue of the adiabatic approximation \cite{p16}.

Before pursuing the derivation of the expression for the 
geometric phase, I must emphasize that a general
cyclic two-component state vector is clearly cyclic in its both
components. Identifying the corresponding function space
${\cal H}_t\oplus{\cal H}_t$ (Note that the $t$-dependence is
only relevant to the inner product structure and the vector
space structure is independent of $t$.) with the space of all
possible initial data, a cyclic two-component state vector
$\Psi^{(q)}(0)$ which by definition satisfies $\Psi^{(q)}(\tau)=
\exp[i\alpha(\tau)]\Psi^{(q)}(0)$, is associated with a `cyclic' 
solution of the Klein-Gordon equation (\ref{k-g-1}) whose
velocity is also cyclic with the same (possibly non-unimodular)
phase and period, i.e., $\Phi(\tau,x^i)=\exp[i\alpha(\tau)]
\Phi(0,x^i)$ and $\dot \Phi(\tau,x^i)=\exp[i\alpha(\tau)]
\dot\Phi(0,x^i)$. This is in contrast with the usual definition
of cyclicity for classical fields \cite{ga-ch,an}. It seems more
reasonable to ascribe the term `cyclic' to a repetition, up
to a scalar multiple, of all initial conditions rather than to
just one of them. In this article I shall use the term {\em cyclic}
in this sense. 

Next, let us consider the two-component formulation of
the Klein-Gordon equation. Suppose for simplicity that
 $E_n^{(q)}$ of Eq.~(\ref{eg-va-eq-q}) is independent
of $q$, i.e., $E_n^{(q)}=E_n$ and that it is non-degenerate.
Then, a direct generalization of the concept of adiabatic
evolution in non-relativistic quantum mechanics suggests
one to use the ansatz:
	\xbe
	\Psi^{(q)}(t)\approx e^{i\alpha_n(t)}\Psi_n^{(q)}[R(t)]\;, 
	\label{rel-ad-ap}
	\xee	
to define the relativistic analog of adiabatic evolution. One
can show, however, that this ansatz leads to a restrictive
notion of adiabatic approximation.
I shall refer to this approximation as the {\em ultra-adiabatic
approximation}. More precisely, I shall use the following 
definition:
	\begin{itemize}
	\item[] {\bf Definition~1:} A two-component state vector
$\Psi^{(q)}(t)$ is said to undergo an {\em exact ultra-adiabatic
evolution} if and only if 
	\xbe
	\Psi^{(q)}(t)=e^{i\alpha_n(t)}\Psi_n^{(q)}[R(t)]\;, 
	\label{ex-rel-ad-ap}
	\xee	
for some $n$ and $\alpha_n$. 
	\end{itemize}
Note that Definition~1 also provides a definition for  {\em ultra-adiabatic
approximation} by replacing Eq.~(\ref{ex-rel-ad-ap}) by
Eq.~(\ref{rel-ad-ap}).

In order to derive the conditions under which the ultra-adiabatic
approximation is valid, one must
substitute Eqs.~(\ref{ex-rel-ad-ap}),  (\ref{eg-va-eq-q}), and
(\ref{eg-ve}) in the Schr\"odinger equation (\ref{sch-eq-q}).
This yields
	\xbea
	\left[ -\dot\alpha_n(1-iqE_n)+q\dot E_n-E_n(1-iqE_n)
	\right]\Phi_n+i(1-iqE_n)\dot \Phi_n&=&0\;,
	\label{I}\\
	\left[ -\dot\alpha_n(1+iqE_n)-q\dot E_n-E_n(1+iqE_n)
	\right]\Phi_n+i(1+iqE_n)\dot \Phi_n&=&0\;.
	\label{II}
	\xeea
Adding both sides of these equations and simplifying the
result, one has
	\xbe
	(\dot\alpha_n+E_n)\Phi_n-i\dot \Phi_n=0\;.
	\label{III}
	\xee
This equation leads directly to the expression for the total phase 
(\ref{e^ia}) with the Berry connection one-form given by
	\xbe
	{\cal A}_n=\frac{i\xbr \Phi_n|\frac{\partial}{\partial R^a}\,
	\Phi_n\xkt}{\xbr \Phi_n| \Phi_n\xkt}\: dR^a=
	\frac{i\xbr \Phi_n|d\Phi_n\xkt}{\xbr \Phi_n| \Phi_n\xkt}\;.
	\label{be-co}
	\xee
Here $R$ denotes the parameters of the system, i.e., the
metric ${\rm g}$, the electromagnetic potential $A$ and the
scalar potential $V$. Moreover I have used the identity 
$\dot \Phi_ndt=(\partial \Phi_n/\partial R^a)\,dR^a=d\Phi_n$.

Furthermore, subtracting Eq.~(\ref{II}) from (\ref{I}) and
using Eq.~(\ref{III}) to simplify the resulting expression,
one finds	
	\xbe
	\dot E_n=0\;.
	\label{rel-condi}
	\xee
This condition which is a direct consequence
of Definition~1 does not have a counterpart in ordinary
non-relativistic quantum mechanics. Its roots may be
sought in the fundamental difference between ordinary
(one-component) Schr\"odinger and Klein-Gordon equations.
One may argue based on physical grounds that the condition 
(\ref{rel-condi}) and consequently the concept of the 
ultra-adiabatic evolution are too restrictive. Indeed it is 
possible to relax this condition
by adopting a more general definition of adiabatic evolution. 
For the moment, however, I shall continue with a further analysis
of the ultra-adiabatic evolutions.

Because Eq.~(\ref{III}) is identical with the one obtained in the
non-relativistic case,  in addition to condition (\ref{rel-condi})
one also has the analog of  Eq.~(\ref{condition}). If $\Phi_n$
turn out to be orthogonal,  the latter reduces to
	\xbe
	\xbr\Phi_m|\dot\Phi_n\xkt=0\;,~~~~~\forall m\neq n\;,	\label{ort-condition}
	\xee
which is the well-known condition for the exactness
of the adiabatic approximation in non-relativistic quantum 
mechanics, \cite{p16}.
Hence, for the cases where $\Phi_n$ are orthogonal, 
the ultra-adiabatic approximation is exact if and  only if 
Eqs.~(\ref{rel-condi}) and (\ref{ort-condition})  are satisfied.

Note that  Eqs.~(\ref{I}) and (\ref{II}) and consequently 
conditions~(\ref{rel-condi}) and (\ref{ort-condition}) are valid
if and only if the ultra-adiabatic approximation is exact.  As one knows
form non-relativistic quantum mechanics, the condition of the
exactness of (ultra-)adiabatic approximation is highly 
restrictive.\footnote{In fact, the (ultra-)adiabatic approximation 
is exact if and only if
the evolving state is stationary \cite{bohm-qm}.} More 
interesting are cases where the (ultra-)adiabatic approximation is valid
only approximately, i.e., cases where instead of (\ref{ex-rel-ad-ap}), 
(\ref{rel-ad-ap}) holds. In this case, Eqs.~(\ref{I}), (\ref{II}), and 
conditions (\ref{rel-condi}) and (\ref{ort-condition}) are required to
be satisfied approximately, namely
	\xbea
	&&\dot E_n\approx 0\;,
	\label{approx-rel-condi}\\
	&&\xbr\Phi_m|\dot\Phi_n\xkt\approx 0\;,~~~~~\forall m\neq n\;.	\label{approx-ort-condition}
	\xeea
More precisely, the ultra-adiabatic approximation is a valid 
approximation if and  only if  (\ref{approx-rel-condi}) and 
(\ref{approx-ort-condition})  are satisfied. The precise meaning
of the $\approx$ in these equations will be clarified
momentarily.

In the above discussion, the condition of time-independence of 
$q$ does not play any significant role in the derivation of Eqs.~(\ref{III})
and (\ref{be-co}). In fact, allowing $q$ to be time-dependent only
changes the term $q\dot E_n$  in Eqs.~(\ref{I}) and (\ref{II}) to
$d(qE_n^{(q)})/dt$. Therefore, up on adding the resulting equations
one still obtains Eq.~(\ref{III}). The only consequences of  using a 
time-dependent $q$ are the emergence of $q$-dependent $E_n$
and $\Phi_n$ and the condition 
	\xbe
	\frac{d}{dt}(qE_n^{(q)})\approx 0\;,
	\label{t-dep-q-condi}
	\xee
which generalizes (\ref{approx-rel-condi}).

There is a particular case in which
$q$ may be time-dependent but $E_n$ and $\Phi_n$ are still
independent of the choice of $q$. This is the case, where the
operator $\hat D_1$ of (\ref{d1}) is zero-th order and it only
involves time-dependent functions. In this case one can choose
$q$ in such a way as to satisfy $\hat D_1=\dot q/q$. This condition
reduces Eq.~(\ref{ge-eg-va-eq}) to the eigenvalue equation
for $\hat D_2$, with eigenvalues $E_n^2$ and eigenvectors
$\Phi_n$. Hence, $E_n$ and $\Phi_n$ are still $q$-independent.
I shall show in section~6 how this apparently very special
case may be realized and used in the study of  spatially homogeneous
(Bianchi) cosmological models.

The appearance of the decomposition parameter $q$ in
(\ref{t-dep-q-condi}) and the fact that this condition has no 
non-relativistic analog suggests that perhaps
the notion of ultra-adiabatic approximation is too limited.
In order to obtain a physically more
appealing concept of adiabatic approximation, one must
consider a more general ansatz than (\ref{rel-ad-ap}).

Consider the general solutions $\Psi$ of the two-component 
Schr\"odinger equation
(\ref{sch-eq-q}) of the form
	\xbe
	\Psi^{(q)}=\sum_n e^{i\alpha_n}\Psi_n^{(q)}\;,
	\label{app-1}
	\xee
where $\alpha_n\in\xC$ and $\Psi_n^{(q)}$ are the eigenvectors 
of the  two-component Hamiltonian (\ref{h-q}). Substituting 
Eq.~(\ref{app-1}) in the Schr\"odinger equation
(\ref{sch-eq-q}) and making use of Eqs.~(\ref{eg-va-eq-q}) and 
(\ref{eg-ve}), one has
	\xbea
	\sum_n e^{i\alpha_n}\left\{ [ (E_n^{(q)}+
	\dot\alpha_n)(1+iqE_n^{(q)})+
	\frac{d}{dt}(qE_n^{(q)})]\Phi_n^{(q)}-
	i(1-iqE_n^{(q)})\dot\Phi_n^{(q)}\right\}&=&0\;,
	\label{app-2-1}\\
	\sum_n e^{i\alpha_n}\left\{ [ (E_n^{(q)}+
	\dot\alpha_n)(1-iqE_n^{(q)})-
	\frac{d}{dt}(qE_n^{(q)})]\Phi_n^{(q)}-
	i(1+iqE_n^{(q)})\dot\Phi_n^{(q)}\right\}&=&0\;.
	\label{app-2-2}
	\xeea
Adding and subtracting both sides of these equations and 
simplifying the result lead to
	\xbea
	&&\sum_ne^{i\alpha_n}[(E_n^{(q)}+\dot\alpha_n)
	\Phi_n^{(q)}-i\dot\Phi_n^{(q)}]=0\;,
	\label{app-2-3}\\
	&&\sum_ne^{i\alpha_n}\left\{[\frac{d}{dt}(qE_n^{(q)})]
	\Phi_n^{(q)}+iqE_n^{(q)}[(E_n^{(q)}+\dot\alpha_n)\Phi_n^{(q)}-
	i\dot\Phi_n^{(q)}]\right\}=0\;.
	\label{app-2-4}
	\xeea

Next assume that $\hat{D}_2$ is a non-degenerate self-adjoint 
operator with a discrete spectrum and $\hat{D}_1=\dot q/q$. Then,
$E_n$ and $\Phi_n$ do not depend on $q$ and $\hat{D}_2\Phi_n=
E_n^2\Phi_n$.  Now, differentiate both sides of  the latter equation 
with respect to time and take their inner
product with $\Phi_m$. Since in this case
$\Phi_n$ are orthogonal, one has the well-known identity 
\cite{berry1984}
	\xbe
	\frac{\xbr\Phi_m|\dot{\hat{D}}_2|\Phi_n\xkt}{E^2_n-E^2_m}=
	\xbr\Phi_m|\dot\Phi_n\xkt\;,~~~~{\rm for~all}~~~m\neq n\;,
	\label{app-3}
	\xee
where $\Phi_n$
and $\Phi_m$  correspond to distinct eigenvalues of $\hat{D}_2$,
i.e., $E_n^2\neq E_m^2$.

Quantum adiabatic approximation is valid if the left hand side of this
equation  which involves the time-derivative of $\hat{D}_2$ can be
neglected, \cite{p16}. This statement provides the true meaning of the
condition (\ref{approx-ort-condition})
	\xbe
	\xbr\Phi_m|\dot\Phi_n\xkt=
	\frac{\xbr\Phi_m|\dot{\hat{D}}_2|\Phi_n\xkt}{E^2_n-E^2_m}
	\approx 0\,,~~~~{\rm for~all}~~m\neq n,
	\label{app-condi}
	\xee
for the case where the above assumptions are valid.
I shall next use this condition to define the notion of {\em adiabatic
evolution} in relativistic scalar quantum mechanics. 

For convenience, I shall use the notation $\Psi_{-n}$ for the 
two-component eigenvector corresponding to the eigenvalue
$E_{-n}:=-E_n$. Since for each pair $(-n,n)$ there is a single
$\Phi_n$, one can write Eqs.~(\ref{app-2-3}) and (\ref{app-2-4})
in the form
	\xbea
	&&\sum_{n\geq 0}\left\{ \left[E_n(e^{i\alpha_n}-e^{i\alpha_{-n}})
	+(\dot\alpha_ne^{i\alpha_n}+\dot\alpha_{-n}e^{i\alpha_{-n}})
	\right]\Phi_n-i(e^{i\alpha_n}+e^{i\alpha_{-n}})\dot\Phi_n\right\}=0,
	\label{4'}\\
	&&\sum_{n\geq 0}\left\{\left[(e^{i\alpha_n}-e^{i\alpha_{-n}})
	\frac{d}{dt}(qE_n)+iqE_n^2(e^{i\alpha_n}+e^{i\alpha_{-n}})+
	iqE_n(\dot\alpha_ne^{i\alpha_n}-\dot\alpha_{-n}e^{i\alpha_{-n}})
	\right]+\right.\xnn\\
	&&~~~~~~~~~\left.qE_n(e^{i\alpha_n}-e^{i\alpha_{-n}})
	\dot\Phi_n\right\}=0.
	\label{5'}
	\xeea
Enforcing condition~(\ref{app-condi}), one can reduce
(\ref{4'}) and (\ref{5'}) to
	\xbea
	E_n(e^{i\alpha_n}-e^{i\alpha_{-n}})
	+(\dot\alpha_ne^{i\alpha_n}+\dot\alpha_{-n}e^{i\alpha_{-n}})
	-(e^{i\alpha_n}+e^{i\alpha_{-n}})\mbox{\large $a$}_n&\approx& 0,
	\label{4''}\\
	(-if_n-\mbox{\large $a$}_n)(e^{i\alpha_n}-e^{i\alpha_{-n}})+E_n
	(e^{i\alpha_n}+e^{i\alpha_{-n}})+\dot\alpha_ne^{i\alpha_n}-
	\dot\alpha_{-n}e^{i\alpha_{-n}}&\approx& 0,
	\label{5''}
	\xeea
where $n\geq 0$ and
	\[ \mbox{\large$a$}_n:=\frac{i\xbr \Phi_n|\dot\Phi_n\xkt}{\xbr 	\Phi_n| \Phi_n\xkt}\,,~~~~~~ f_n:=\frac{\frac{d}{dt}(qE_n)}{qE_n}
	=\frac{d}{dt}\ln(qE_n)\;.\]
Adding and subtracting both sides of (\ref{4''}) and (\ref{5''})
and assuming that $e^{i\alpha_n}$ is not negligibly small,
one finds
	\xbea
	-if_n(1-e^{-i(\alpha_n-\alpha_{-n})})+2(E_n+\dot\alpha_n-
	\mbox{\large $a$}_n)&\approx&0\;,
	\label{6'}\\
	-if_n(e^{i(\alpha_n-\alpha_{-n})}-1)+2(E_n-\dot\alpha_{-n}+
	\mbox{\large $a$}_n)&\approx&0\;.
	\label{7'}
	\xeea
Next, define $\eta_n^-:=\alpha_n-\alpha_{-n}$, add both sides
of (\ref{6'}) and (\ref{7'}), and simplify the result. This leads to
	\xbe
	\dot\eta_n^-+f_n\sin\eta_n^-+2E_n\approx 0\;.
	\label{8'}
	\xee
Introducing $\eta_n^+:=\alpha_n+\alpha_{-n}$ and
using (\ref{8'}), one can then express (\ref{6'}) in the form
	\xbe
	\dot\eta_n^+-if_n(1-\cos\eta_n^-)-2\mbox{\large $a$}_n\approx 0\;,
	\label{9'}
	\xee
Hence in view of the definition $\eta_n^\pm:=\alpha_n\pm
\alpha_{-n}$, one has
	\xbea
	\alpha_{\pm n}&=&\frac{1}{2}(\eta_n^+\pm\eta_n^-)\approx
	\gamma_n+\delta_{\pm n}\;,~~~~~~{\forall n\geq 0},
	\label{10'}\\
	\gamma_n&:=&\int_0^t\mbox{\large $a$}_n(t')dt'=
	\int_{R(0)}^{R(t)}{\cal A}_n[R]\;,
	\label{11'}\\
	\delta_{\pm n}&:=&\frac{i}{2}\int_0^tf_n(t')[1-\cos\eta_n(t')]dt'
	\pm\frac{\eta_n(t)}{2}\;,
	\label{12'}
	\xeea
where I have used Eq.~(\ref{be-co}) and $\eta_n$
is the solution of 
	\xbe
	\dot\eta_n+f_n\sin\eta_n+2E_n= 0
	\;,~~~~{\rm with}~~~~\eta_n(0)=0\;.
	\label{8''}
	\xee

As seen from (\ref{10'})-(\ref{12'}), the part $\gamma_n$ of 
$\alpha_{\pm n}$ which is independent of $E_n$ has the same 
form as the geometric phase angle of the non-relativistic
quantum mechanics. In contrast, the part $\delta_{\pm n}$ of
$\alpha_{\pm n}$ which does depend on $E_{\pm n}$ and
plays the role of the dynamical phase angle, has a different
expression form its non-relativistic counterpart. For the case 
of an ultra-adiabatic evolution
where condition (\ref{t-dep-q-condi}) is satisfied,
$f_n\approx 0$ and 
	\[\delta_{\pm n}=\mp\int_0^t E_n(t') dt'
	=-\int_0^t E_{\pm n}(t') dt'\;.\]
This is identical with the expression for the non-relativistic
adiabatic dynamical phase. 

The above analysis shows that taking (\ref{app-condi}) as the
defining condition for the {\em adiabatic approximation}, one 
obtains the same expression for the geometric phase as
in the ultra-adiabatic case. This condition modifies the
expression for the dynamical phase. In fact, the dynamical
phase angle splits into a pair $(\delta_{-n},\delta_{n})$ of
dynamical angles. The latter is a consequence of the violation
of the ultra-adiabaticity condition (\ref{app-condi}).

The relativistic adiabatic approximation outlined in the 
preceding paragraphs corresponds
to the following definition of relativistic adiabatic evolution
	\begin{itemize}
	\item[]{\bf Definition 2:} A two-component state vector
$\Psi^{(q)}(t)$ is said to undergo an {\em exact adiabatic
evolution} if and only if 
	\xbe
	\Psi^{(q)}(t)= e^{i\alpha_n(t)}\Psi_n^{(q)}[R(t)]+
	e^{i\alpha_{-n}(t)}\Psi_{-n}^{(q)}[R(t)]\;, 
	\label{ex-rel-ad-ap-3}
	\xee	
for some $n$ and $\alpha_{\pm n}$. 
	\end{itemize}
The {\em approximate adiabatic approximation} corresponds
to the case where (\ref{ex-rel-ad-ap-3}) is approximately valid,
i.e.,
	\xbe
	\Psi^{(q)}(t)\approx e^{i\alpha_n(t)}\Psi_n^{(q)}[R(t)]+
	e^{i\alpha_{-n}(t)}\Psi_{-n}^{(q)}[R(t)]\;.
	\label{ex-rel-ad-ap-2}
	\xee
This approximation is valid if and only if $\xbr\Phi_m|
\dot\Phi_n\xkt \approx 0$ for all $m\neq n$.	
Definition~2 provides a suitable definition for an adiabatic
evolution in relativistic (scalar) quantum mechanics. In
particular, it ensures that for a cyclic change of the
parameters of the system, the one-component Klein-Gordon 
field and its time-derivative have cyclic evolutions. The difference
between the ultra-adiabatic and adiabatic evolutions is that
for a cyclic ultra-adiabatic evolution the (possibly non-unimodular)
phases of the one-component field and its time-derivative
are required to be equal, whereas in a cyclic adiabatic evolution
these phases are generally different. 

Since the defining condition for the relativistic and 
non-relativistic adiabatic evolution are identical, one 
can use the well-known results of non-relativistic 
quantum mechanics to generalize the above results to 
the case where $E_n$ is degenerate.

An important aspect of the above derivation of the geometric
phase is that it does not use the particular form of  an inner
product on ${\cal H}_t\oplus{\cal H}_t$, i.e., the Hermitian
matrix $h$ of (\ref{he-in-pr}). It only uses the inner product on
${\cal H}_t$. One might argue that based on the known 
features of the non-relativistic case, the independence of the
geometric phase from the inner product is quite natural and
thus not particularly important. A review of the existing literature 
\cite{an-ma,co-pi} shows, however, that in the previously studied
examples a great deal of effort was made to define an inner
product on the space of solutions before the problem of the
geometric phase could be addressed. The construction of
such an inner product  is a highly technical problem and a
satisfactory solution for arbitrary spacetimes is not known. The results
of this section indicates that indeed one does not need to construct
an inner product on the space of solutions. What is needed is
the $L^2$  inner product on ${\cal H}_t $ which is naturally given
by the induced spatial metric. In this way, one can conveniently
avoid the difficult problem of  constructing an inner product on
the space of solutions and carry on with the analysis of
the adiabatic geometric phase.

In the following sections the practical advantages of the two-component
formulation are demonstrated for three physical examples.
	
\section{Rotating Magnetic Field in Minkowski Background}

Consider the geometric phase induced on a Klein-Gordon
field in a Minkowski background due to a rotating magnetic
field. This problem was originally studied by Anandan and Mazur 
\cite{an-ma} using the one-component formalism.

In this case, in a global cartesian coordinate system, one has 
$g_{00}=-1,~ g_{ij}=\delta_{ij}$, $g_{0i}=V=0$,  and ${\cal H}_t
=L^2(\xR^3)$. Following \cite{an-ma}, let us first consider the
case of a constant magnetic field along the $x^3$-axis. Then in
the symmetric gauge, one has $A_0=A_3=0$, $A_1=-Bx^2/2$,
and $A_2=Bx^1/2$.  Substituting these equations in Eqs.~(\ref{d1})
and (\ref{d2}), one finds $\hat D_1=0$ and
	\xbe
	\hat D_2=-\nabla^2-ieB\frac{\partial}{\partial\varphi}+
	\frac{e^2B^2}{4}\:\rho^2+\mu^2\;,
	\label{d2-landau}
	\xee
where $\nabla^2$ is the Laplacian and $(\rho,\varphi,x^3)$
are cylindrical coordinates in $\xR^3$. Clearly
$\hat D_2$ is self-adjoint. Therefore,  Eq.~(\ref{ge-eg-va-eq})
reduces to the eigenvalue equation for $\hat D_2$, namely
$\Phi_n$ are orthogonal eigenvectors of $\hat D_2$ with
eigenvalue $E_n^2$. Taking $q=i$ in  Eq.~(\ref{u-v-q}) and
choosing the inner product (\ref{he-in-pr-sp}), the Hamiltonian
$H^{(i)}$ of (\ref{sch-eq-q}) is also self-adjoint. 

The situation is quite similar to the non-relativistic Landau
level problem. Clearly, $\Phi_n$ are infinitely degenerate. They
are given by
	\xbe
	\Phi_n^{(p,m)}=N_ne^{ipx^3}e^{im\varphi}\:
	\chi_{nmp}(\rho)\;,
	\label{un-landau}
	\xee
where $p\in\xR$,  $m=0,1,2,\cdots$ label the vectors
within the degeneracy subspace ${\cal H}_n$, $\chi_{n
mp}$ are orthogonal solutions of
	\xbea
	\left[ \frac{d^2}{d\rho^2}+\frac{1}{\rho}\frac{d}{d\rho}\,
	+(k^2-\frac{m^2}{\rho^2}-
	\lambda^2\,\rho^2)\right] \chi_{nmp}(\rho)&=&0\,,
	\label{chi}\\
	k^2\::=\:E_n^2-(p^2+\mu^2+emB)\;,~~~~\lambda&
	:=&\frac{eB}{2}\xnn
	\xeea
and $N_n$ are normalization constants chosen in such a
way as to ensure
	\xbe
	\xbr \Phi_{\tilde n}^{(\tilde p,\tilde m)}|\Phi_n^{(pm)} \xkt=
	\delta(\tilde n,n) \:\delta(\tilde m ,m)\:\delta(\tilde p,p)\;.
	\label{normalization}
	\xee
Here $\delta(~,~)$ denotes a Kronecker or a Dirac delta
function depending on whether the arguments are discrete
or continuous, respectively.

In order to solve the eigenvalue problem for the
rotating magnetic field, one can easily use the unitary
transformations \cite{an-ma,bohm-qm}
	\xbe
	{\cal U}(\theta,\varphi)=e^{-i\varphi \hat J_3} e^{-i\theta 
	\hat J_2}e^{i\varphi \hat J_3}\;,
	\label{u}
	\xee
relating the eigenvectors $\Phi_n$ of $\hat D_2$
to those corresponding to the constant magnetic field
(\ref{un-landau}). In Eq.~(\ref{u}), $\theta$ and $\varphi$
are azimuthal and polar angles in spherical coordinates
and $\hat J_i$ are angular momentum operators (generators
of $SO(3)$) acting on the Hilbert space $L^2(\xR^3)$.
${\cal U}(\theta,\varphi)$ are well-defined everywhere except
along the negative $x^3$-axis which can be excluded by
assuming that $\vec B(t)=(B,\theta(t),\varphi(t))$ does not cross
this axis.  Otherwise, one may choose another coordinate
frame and remedy the problem by performing appropriate
gauge transformations as described in Ref.~\cite{bohm-qm}
for the non-relativistic case. Clearly,
	\xbea
	\hat D_2[\vec B(t)]&=& {\cal U}(\theta(t),\varphi(t))\:
	\hat D_2[\vec B=B\hat x^3]\:{\cal U}^\dagger(\theta(t),
	\varphi(t))\;,\xnn\\
	\Phi_n[\vec B(t)]&=&{\cal U}(\theta(t),\varphi(t))\:\Phi_n[\vec B=
	B\hat x^3]\;,\xnn\\
	E_n[\vec B(t)]&=&E_n[\vec B=B\hat x^3]\:=\:
	{\rm constant}\,. \xnn
	\xeea
The latter relation which implies $\dot E_n=0$ indicates that
an adiabatic evolution of this system is, in fact, ultra-adiabatic.

As noted in Ref.~\cite{an-ma}, the presence of the degeneracy
leads to non-Abelian geometric phases (\ref{e^ia}) defined
by the connection one-form ${\cal A}_n$, (\ref{connection}).
 The components of ${\cal A}_n$ are given by the non-Abelian
generalization of (\ref{be-co}), namely
	\xbe
	A^{IJ}_n=i\xbr \Phi^{(I)}_n|d \Phi^{(J)}_n\xkt\;,
	\label{non-abelian-connection}
	\xee
and are independent of the choice of the matrix $h$ of
(\ref{he-in-pr}). In Eq.~(\ref{non-abelian-connection}),
$I:=(p,m)$ and $J:=(p',m')$, and use is made of
(\ref{normalization}). I shall not be elaborating on this
problem any further since the specific results are exactly the
same as the ones reported in Ref.~\cite{an-ma}. It is
however worth mentioning that each $\Phi_n$ defines
a pair of orthonormal two-component eigenvectors 
$\Psi_{\pm n}^{(i)}$ corresponding to the choices $\pm E_n$
for each eigenvalue $E_n^2$ of $\hat D_2$. Hence in
this case the two-component formalism reproduces the
results of \cite{an-ma} which were obtained using a more
subtle method of taking square root of the second order
Klein-Gordon operator and projecting onto the spaces of
negative and positive energy (frequency) solutions of the
Klein-Gordon equation.

\section{Rotating Cosmic String}

In Ref.~\cite{co-pi}, the authors study the geometric (or rather
topological) phases induced on a Klein-Gordon field due to a
rotating cosmic string. In this section, I shall outline a solution to
this problem using the two-component formalism. 

The local coordinate expression for the metric corresponding
to a rotating cosmic string with angular momentum $j$ and
linear mass density $d$ is \cite{co-pi}:
	\xbe
	{\rm g}=\left(
	\begin{array}{cccc}
	-1&0&-4j&0\\
	0&1&0&0\\
	-4j&0&(\alpha\rho)^2-(4j)^2&0\\
	0&0&0&1\end{array}\right)\;,
	\label{metric-cs}
	\xee
where $(x^\mu)=(t,\rho,\varphi,z)$ and $(\rho,\varphi,z)$ are
cylindrical coordinates on the spatial hypersurface $\Sigma_t$
and $\alpha:=1-4d$. $\Sigma_t$ corresponds to a cone with
a deficit angle $\beta=8\pi d=2\pi (1-\alpha)$. 

Note that for $\rho\leq 4j/\alpha$, $\partial/\partial\varphi$
becomes timelike. This leads to the existence of closed
timelike curves. This region can be ignored by imposing
appropriate boundary conditions on the fields, i.e.,
$\Phi=0$ for $\rho\leq 4j/\alpha$.

Performing the necessary calculations, one finds the following
expressions for the operators $\hat D_1$ and $\hat D_2$ of
Eqs.~(\ref{d1}) and (\ref{d2}):
	\xbea
	\hat D_1&=&\frac{8j}{(\alpha\rho)^2-(4j)^2}
	\:\frac{\partial}{\partial\varphi}\;,
	\label{d1-cs}\\
	\hat D_2&=&(\frac{-1}{1-(\frac{4j}{\alpha\rho})^2})
	\left[ \frac{\partial^2}{\partial\rho^2}+\frac{1}{\rho}\:
	\frac{\partial}{\partial\rho}+\frac{1}{(\alpha\rho)^2}\:
	\frac{\partial^2}{\partial\varphi^2}+\frac{\partial^2}{
	\partial z^2}-\mu^2\right]\;.
	\label{d2-cs}
	\xeea
Therefore the conditions for the self-adjointness of the
Hamiltonian $H^{(q)}$ of (\ref{sch-eq-q}) cannot be met.
Let us proceed, however, with considering the eigenvectors
of  $\Psi_n^{(q)}$ of $H^{(q)}$, (\ref{eg-va-eq-q}). For the
metric (\ref{metric-cs}), Eq.~(\ref{ge-eg-va-eq-2}) takes the
form:
	\xbe
	\left\{  \frac{\partial^2}{\partial\rho^2}+\frac{1}{\rho}\:
	\frac{\partial}{\partial\rho}+\frac{1}{(\alpha\rho)^2}\:
	\frac{\partial^2}{\partial\varphi^2}+
	\frac{i8jE_n}{(\alpha\rho)^2}\: \frac{\partial}{\partial
	\varphi}+
	\frac{\partial^2}{\partial z^2}-\mu^2+
	[1-(\frac{4j}{\alpha\rho})^2]E_n^2
	\right\}\Phi_n
	=0\;.
	\label{ge-ei-va-eq-cs}
	\xee
In view of an observation made in Ref.~\cite{ge-ja} and
used in \cite{co-pi}, let us write $\Phi_n$ in the form $\Phi_n
=\exp(i\zeta\varphi)\phi_n$. Substituting this equation in
(\ref{ge-ei-va-eq-cs}), one finds that for $\zeta=-4jE_n$,
$\phi_n$ satisfies:
	\xbe
	\left\{  \frac{\partial^2}{\partial\rho^2}+\frac{1}{\rho}\:
	\frac{\partial}{\partial\rho}+\frac{1}{(\alpha\rho)^2}\:
	\frac{\partial^2}{\partial\varphi^2}+\frac{\partial^2}{
	\partial z^2}-\mu^2+E_n^2\right\} \phi_n=0\;.
	\label{phi-n}
	\xee
Eq.~(\ref{phi-n}) may be obtained from (\ref{ge-ei-va-eq-cs})
by setting $j=0$ and replacing $\Phi_n$ by $\phi_n$.
Hence $\phi_n$ determine the eigenvectors of the
Hamiltonian for a non-rotating string of the same mass
density. In this case $\hat D_1$ vanishes and $\hat D_2$
becomes self-adjoint.  Therefore, $\phi_n$ are orthogonal
eigenvectors of $\hat D_2$, with $j=0$. Choosing $q=i$ and
adopting the inner product (\ref{he-in-pr-sp}), the corresponding
Hamiltonian becomes also self-adjoint.

In  fact, it is not  difficult to show that the solutions of
Eq.~(\ref{phi-n}) are of the form:
	\xbe
	\phi_n=N_n e^{ipz}e^{im\varphi}\:J_\nu(k\rho)\;,
	\label{phi-n=}
	\xee
where $N_n$ are appropriate normalization constants,
$J_\nu$ are Bessel functions,  and
	\[ k:=\sqrt{\:E_n^2-(p^2+
	\mu^2)}\;,~~~~~ \nu:=m/\alpha\;.\]

The orthogonality property of $\phi_n$ carries over to 
$\Phi_n$ since the measure of the integration on $\Sigma_t$
is independent of $j$. This is because of the identity:
	\[ \det[{\rm g}]=-\det[^{(3)}{\rm g}]\;,\]
which holds for any metric with the lapse function $N=1$,
\cite{wald-gr}.

The situation is analogous to the case of a rotating
magnetic field. However, in this case the
Klein-Gordon field acquires an Aharonov-Bohm type
phase which is topological in nature.  
As Berry describes in his (by now classic) article
\cite{berry1984}, the Aharonov-Bohm phase may be viewed
as a particular case of a geometric phase. This is done,
for the original Aharonov-Bohm system of an electron
encircling a confined magnetic flux line, by considering
the electron to be localized in a box which is then carried
around the flux line. Thus the time-dependence of the
system is introduced by choosing a coordinate system
centered  inside the box. This leads to geometric phases
for the energy eigenfunctions. The same result is then
applied to the electron wave packet, only because
the geometric phase is independent of the energy
eigenvalues, i.e., all the energy eigenvectors and
therefore any linear combinations of them, in particular
the one forming the localized electron wave packet,
acquire the same geometric phase which  is then shown
to be the same as the one discovered by
Aharonov and Bohm \cite{ah-bo}.

Ref.~\cite{co-pi} uses the analogy between the system
of rotating cosmic string and that of Aharonov and Bohm
to obtain the corresponding geometric phases. This is however
not quite justified for arbitrary energy eigenfunctions since as
shown below and also in \cite{co-pi}, unlike the
Aharonov-Bohm system, the induced phase in this case does
depend on the energy eigenvalue. Consequently an arbitrary
localized Klein-Gordon field which is a superposition of different
energy eigenfunctions will not be cyclic.  Berry's argument
therefore applies only to those localized field configurations
which are energy eigenfunctions. Such eigenfunctions  do
actually exist. Particular examples can be constructed by virtue
of the infinite degeneracy arising from the axial symmetry
of the problem. This is demonstrated in Appendix~A.

Therefore, we can proceed with using
the analogy with Berry's treatment of the Aharonov-Bohm
phase \cite{berry1984} to derive the geometric phase in the
framework of the two-component formalism. This is done by
changing to a frame centered in a box which circulates
around the string at a distance larger than $4j/\alpha$.
If $R^i$ are coordinates of the center of the box and $x^{'i}$
are the coordinates centered at $R=(R^i)$, then the
eigenfunctions are of the form: $\Phi_n(x')=\Phi_n(x-R)$.
Substituting this expression in the non-Abelian version of
(\ref{be-co}), one finds
	\xbea
	{\cal A}_n^{IJ}&=&i\xbr \Phi_n^I(x-R)|\frac{
	\partial}{\partial R^i}|\Phi_n^J(x-R)\xkt\: dR^i\;,
	\xnn\\
	&=&i\int _\Sigma d\Omega \:\phi^{*I}_n(x-R) \left[
	-4iE_nj \phi^J_n(x-R)\, dR^2 +\frac{\partial}{
	\partial R^i}\phi_n^J(x-R)\,dR^i\right] \;,\xnn\\
	&=&4jE_n\delta_{IJ}\:dR^2\;,
	\label{be-co-cs}
	\xeea
where $d\Omega=\alpha\rho\,d\rho\, d\varphi\, dz$,
$I$ and $J$ stand for possible degeneracy labels 
corresponding to localized eigenfunctions, $R^2$ is the
polar angle associated with the center of the box, and
$\phi_n$ are assumed to be normalized.  For a curve $C$
with winding number $N_C$, the geometric phase `angle'
is given by
	\xbe
	\gamma_n=N_C\int_0^{2\pi}{\cal A}_n=
	8\pi jE_n\,N_C\;,
	\label{ge-ph-an}
	\xee
where the labels $I,J$ and $\delta_{IJ}$ have been
suppressed for convenience.
This is identical with the result of Ref.~\cite{co-pi}.
Note however that here I have not been concerned
with the consideration of the difficult problem of the
choice of an inner product for the space of the
solutions of the Klein-Gordon equation (${\cal H}_t\oplus
{\cal H}_t$), such as the one proposed by Ashtekar and
Mangon \cite{as-ma} and apparently `used' by Corichi
and Pierri in Ref.~\cite{co-pi}. In fact, as I have shown in
section~3, the geometric phase is independent of  the
particular choice of such an inner product. This is also
implicit in the Corichi and Pierri's derivation of
the geometric phase in \cite{co-pi}. Although they
discuss the Ashtekar-Mangon scheme in some detail,
the final derivation does not use the particular form
of the inner product.

It is also worth mentioning that although the eigenvalues
$E_n$ may be degenerate, the corresponding geometric
phase is still Abelian. 

\section{Spatially Homogeneous Cosmological Models}
Consider  Klein-Gordon fields in a spatially homogeneous (Bianchi)
cosmological background associated with a Lie group
$G$, i.e., $M=\xR\times G$. In a synchronous invariant
basis the spacetime metric g is given by its spatial
components $g_{ab}$:
	\xbe
	ds^2=g_{\alpha\beta}\omega^\alpha\omega^\beta=
	-dt^2+g_{ab}\omega^a\omega^b\;,
	\label{ds2}
	\xee
where $\omega^a$ are the left invariant one-forms
and $g_{ab}=g_{ab}(t)$.  Throughout this article I use the
conventions of Ref.~\cite{ra-sh}.

The first step in the study of the phenomenon of geometric
phase due to a spatially homogeneous cosmological background
is to compute the operators $\hat D_1$ and $\hat D_2$ of
Eqs.~(\ref{d1}) and (\ref{d2}) in the invariant basis.  It is not
difficult to see that with some care these equations are valid
for arbitrary basis. One must only replace the coordinate
labels ($\mu,\nu,\cdots, i,j,\cdots$) with the basis (in this case
invariant basis) labels $(\alpha,\beta,\cdots,a,b,\cdots)$,
and interpret $\partial_a$ as the action of the operators
$\hat X_a$ associated with the dual vector fields to $\omega^a$.
This leads to:
	\xbea
	\hat D_1&=&g^{ab}\Gamma_{ab}^0\;,
	\label{d1-bc}\\
	\hat D_2&=& -\Delta_t+\mu^2\;,
	\label{d2-bc}\\
	\Delta_t&:=&g^{ab}\nabla_a\nabla_b\:=\:
	g^{ab}\hat X_a\hat X_b-\Gamma_{ab}^c\hat X_c\;,
	\label{laplacian-bc}
	\xeea
where $\Delta_t$ is the Laplacian on $\Sigma_t$, $\nabla_a$
are the covariant derivatives corresponding to the Levi Civita
connection, and \cite{ra-sh}:
	\xbe
	\Gamma_{\alpha\beta}^\gamma:=\frac{1}{2}\,
	g^{\gamma\delta}(g_{\delta\alpha,\beta}+
	g_{\beta\delta,\alpha}-g_{\alpha\beta,\delta}+
	g_{\epsilon\alpha}C^{\epsilon}_{\delta\beta}+
	g_{\epsilon\beta}C_{\delta\alpha}^\epsilon)-
	\frac{1}{2}\:C_{\alpha\beta}^\gamma\;.
	\label{gamma}
	\xee
Here $g_{\alpha\beta,\gamma}=\hat X_\gamma g_{
\alpha\beta}$ and  $C_{\alpha\beta}^\gamma$ are structure
constants:
	\xbe
	\left[\hat X_\alpha,\hat X_\beta\right]=-
	C_{\alpha\beta}^\gamma\, \hat X_\gamma\;,
	\label{lie-al}
	\xee
with $\hat X_0:=\partial/\partial t$. In view of the latter
equality, the structure constants with a time label vanish.
This simplifies the calculations of $\Gamma$'s. The only
nonvanishing ones are
	\xbea
	\Gamma_{ab}^0&=&\frac{1}{2}\:\dot g_{ab}\;,
	\label{gamma-ab0}\\
	\Gamma_{ab}^c&=&\frac{1}{2}\: g^{cd}(g_{ea}C^e_{db}+
	g_{eb}C^e_{da})-\frac{1}{2}\:C_{ab}^c\;.
	\label{gamma-abc}
	\xeea
In view of these relations, the expression for $\hat D_1$
and $\hat D_2$ may be further simplified:
	\xbea
	\hat D_1&=& \frac{\partial}{\partial t}\ln\sqrt{g}\;,
	\label{d1-bc-2}\\
	\hat D_2&=& -\Delta_t+\mu^2\:=\:-(
	g^{ab}\hat X_a\hat X_b-C_{ab}^bg^{ac}\hat X_c)+\mu^2\;,
	\label{d2-bc-2}
	\xeea
where $g$ is the determinant of $(g_{ab})$. Note that for
a unimodular, in particular semisimple, group $C_{ab}^b=0$,
and the second term on the right hand side of  (\ref{d2-bc-2})
vanishes. The corresponding Bianchi models are knows as
Class A models.

As seen from Eq.~(\ref{d1-bc-2}), $\hat D_1$ acts by
multiplication by a time-dependent function. Therefore,
choosing $q=i\sqrt{g}$ reduces Eq.~(\ref{ge-eg-va-eq})
to the eigenvalue equation:
	\xbe
	(\hat D_2-E_n^2)\Phi_n=-(\Delta_t+E_n^2-\mu^2)
	\Phi_n=0\;,
	\label{eg-va-eq-la}
	\xee
for the operator $\hat D_2$ which being essentially the
Laplacian over $\Sigma_t$, is self-adjoint. This
guarantees the orthogonality of $\Phi_n$ and the reality of
$E_n^2$. Furthermore, since $q$ is imaginary, the
Hamiltonian $H^{(q)}$ with the choice of (\ref{he-in-pr-sp})
for the inner product is also self-adjoint. 

The analysis of the $\Phi_n$ is equivalent to the study of
the eigenvectors of the Laplacian over a three-dimensional
group manifold $\Sigma_t$. The general problem is the subject
of the investigation in spectral geometry which is beyond the
scope of the present article. However, let us  recall some
well-known facts about spectral properties of the Laplacian
$\Delta$ for an arbitrary finite-dimensional Riemannian
manifold $\Sigma$ without boundary.

The following results are valid for the case where
$\Sigma$ is compact or the eigenfunctions are required to
have a compact support\footnote{This is equivalent with the
case where $\Sigma$ has a boundary $\partial\Sigma$, over
which the eigenfunctions vanish.}
\cite{ga-hu-la}:
	\begin{itemize}
	\item[1.] The spectrum of $\Delta$ is an infinite discrete
	subset of  non-negative real numbers.
	\item[2.] The eigenvalues are either non-degenerate or
	finitely degenerate. 
	\item[3.] There is an orthonormal set of eigenfunctions
	which form a basis for $L^2(\Sigma)$.
	\item[4.] If $\Sigma$ is compact,  then the first eigenvalue
	is zero which is non-degenerate with the eigenspace
	given by the set of constant functions, i.e., $\xC$. If 
	$\Sigma$ is not compact but the eigenfunctions are
	required to have a compact support, then the first
	eigenvalue is positive.
	\end{itemize}

In view of Eq.~(\ref{eg-va-eq-la}) and the results concerning
the spectral properties of  the Laplacian $\Delta_t$, one has
some restrictions on the energy eigenvalues $E_n$. The
non-negativity of the eigenvalues $\mu^2-E_n^2$ of $\Delta_t$
is a necessary condition for the validity of the first quantized
theory. A violation of this condition is equivalent to the Klein
paradox \cite{holstein}. In fact, similar conditions exist for both
rotating magnetic field and rotating cosmic string problems.
In these cases, it indicates that the variables $k$ appearing
in the analysis of these systems must be imaginary.

Another piece of useful information about the spatially 
homogeneous cosmological models is that $\Sigma_t$ is
a group manifold, i.e., topologically it is identical to a 
three-dimensional Lie
group $G$. Therefore, one can use the canonical
(left and right) invariant metric $(\delta_{ab})$ on $G$, to
define the corresponding Laplacian $\Delta_0$ and use an
orthonormal eigenbasis $f_n$ of $\Delta_0$ to span
$L^2(\Sigma_t)={\cal H}_t$, i.e., express $\Phi_n$ as a linear
combination of $f_n$.  Perhaps more importantly,
 using the fact that (up to a multiple of $i=\sqrt{-1}$) the
invariant vector fields $\hat X_a$ yield a representation of
the generators $L_a$ of $G$, with $L^2(\Sigma_t)$ being the
representation space, one can view the Laplacian $\Delta_t$
as (a representation of ) of an element of the enveloping algebra
of the Lie algebra of $G$. Therefore, $\Delta_t$ commutes with
any Casimir operator ${\cal C}_\lambda$ and consequently
shares a set of simultaneous eigenvectors with
${\cal C}_\lambda$. This in turn suggests one to specialize
to particular subrepresentations with definite
${\cal C}_\lambda$. In particular for compact groups, this
leads to a reduction of the problem to a collection of
finite-dimensional ones.\footnote{Here I mean a
finite-dimensional Hilbert space.}

In the remainder of this article I shall  try to employ these
considerations to investigate some specific models.

\subsection{Bianchi Type I}

In this case $G$ is Abelian, therefore $X_a$ are themselves
Casimir
operators and the eigenfunctions of  $\Delta_t$, i.e., $\Phi_n$,
are independent of $t$. Hence the Berry connection one-form
(\ref{be-co}) vanishes identically and the geometric phase is
trivial.\footnote{By a trivial geometric phase, I mean a zero
geometrical phase angle.}

\subsection{Bianchi Type IX}

In this case $G=SU(2)$. The total angular momentum operator
$\hat J^2=\sum_a \hat J_a^2$ is a Casimir operator. Therefore,
I shall consider the subspaces ${\cal H}_j$ of ${\cal H}_t=
L^2(S^3_t)$ of definite angular momentum $j$. The left-invariant
vector fields $\hat X_a$ are given by $\hat X_a=i\hat J_a$, in
terms of which Eq.~(\ref{lie-al}), with $C_{ab}^c=\epsilon_{abc}$,
is written in the familiar form:
	\xbe
	\left[ \hat J_a,\hat J_b\right]=i\epsilon_{abc}\hat J_c\;,
	\label{lie-al-su2}
	\xee
with $\epsilon_{abc}$ denoting the totally antisymmetric Levi
Civita symbol and $\epsilon_{123}=1$. 

Eq.~(\ref{eg-va-eq-la}) takes the form:
	\xbe
	(\hat{H'}+k^2_n)\Phi_n=0\;,
	\label{eg-va-eq-tilde}
	\xee
where $\hat{H'}$ is an induced Hamiltonian defined by
	\xbe
	\hat H':=	g^{ab}(t) \hat J_a\hat J_b\;,
	\label{tilde-h}
	\xee
and $k^2_n:=E_n^2-\mu^2$. Therefore, the problem of the
computation of the geometric phase is identical with
that of the non-relativistic quantum mechanical system
whose Hamiltonian is given by (\ref{tilde-h}). In
particular, for the mixmaster spacetime, i.e., for $g_{ab}$
diagonal, the problem is identical with the quantum mechanical
problem of a non-relativistic asymmetric
rotor, \cite{hu}.

Another well-known non-relativistic
quantum mechanical effect which is described by
a Hamiltonian of the form (\ref{tilde-h}) is the quadratic
interaction of a spin with a variable electric field $(E_a)$.
The interaction potential is the Stark Hamiltonian: $\hat H_S
=\epsilon (\sum_a E_a\hat J_a)^2$. The phenomenon of the
geometric phase for the Stark Hamiltonian for spin $j=3/2$,
which involves Kramers degeneracy \cite{messiyah}, was
first considered by Mead \cite{mead}. Subsequently,
Avron,~et~al~\cite{av-sa-se-si,av2} conducted a thorough
investigation of the traceless quadrupole Hamiltonians of the
form (\ref{tilde-h}). 

The condition on the trace of the Hamiltonian is
physically irrelevant, since the addition of any
multiple of the identity operator to the Hamiltonian
does not have any physical consequences. In general,
one can express the Hamiltonian (\ref{tilde-h}) in the
form $\hat{H'}=\hat{\tilde H'}+\hat H'_0$, where
$\hat{\tilde H'}:=Tr(g^{ab})\hat J^2/3$,
	\xbe
	\hat H'_0[R]:=\sum_{A=1}^5 R^A\: \hat e_A\;,	\label{h'0}
	\xee
is the traceless part of the Hamiltonian, and
	\xbea
	\hat e_1&:=&J_3^2-\frac{1}{3}\: \hat J^2\;,~~~~~~~~
	\hat e_2\::=\: \frac{1}{\sqrt{3}}\: \{\hat J_1,\hat J_2\}\;,
	\xnn\\
	\hat e_3&:=&\frac{1}{\sqrt{3}}\: \{\hat J_2,\hat J_3\}
	\;,~~~~~~~~\hat e_4\::=\: \frac{1}{\sqrt{3}}\:(\hat J_1^2
	-\hat J_2^2)\;,
	\label{e's}\\
	\hat e_5&:=& \frac{1}{\sqrt{3}}\:\{\hat J_1,\hat J_2\}\;,
	\xnn\\
	R^1&:=&g^{33}-\frac{1}{2}\:(g^{11}+g^{22})
	\;,~~~~~~~~R^2\::=\:\sqrt{3}\: g^{13}\;,\xnn\\
	R^3&:=&\sqrt{3}\: g^{23}\;,~~~~~~~~
	R^4\::=\: \frac{\sqrt{3}}{2}\:(g^{11}-g^{22})\;,
	\label{R's}\\
	R^5&:=&\sqrt{3}\: g^{12}\;.\xnn
	\xeea
As shown in Refs.~\cite{av-sa-se-si,av2} the space
${\cal M}'$ of all traceless Hamiltonians of  the form
(\ref{h'0}) is a five-dimensional real vector space. The
traceless operators $\hat e_A$ form an orthonormal
\footnote{Orthonormality is defined by the inner product
$\xbr \hat A,\hat B\xkt:=3\:Tr(\hat A\hat B)/2$.} basis for
${\cal M}'$. Removing the point $(R^A=0)$ from ${\cal M}'$,
to avoid the collapse of all eigenvalues, one has therefore
the space $\xR^5-\{ 0\}$ as the parameter space. The
situation is analogous to Berry's original example of a
magnetic dipole in a changing magnetic field,
\cite{berry1984}. Again, a rescaling of the Hamiltonian
by a non-zero function of $R^A$ does not change the
geometric phase. Thus the relevant parameter space is
${\cal M}=S^4$. Incidentally, the point corresponding
to $0\in\xR^5$ which is to be excluded, corresponds to
the class of Friedmann-Robertson-Walker models.

Clearly a curve $C:[0,T]\to S^4$ may be defined by the
action of the group $SO(5)$ which acts transitively
on $S^4$. Therefore the time-dependence of the
Hamiltonian may be realized by an action of the
group $SO(5)$ on a fixed Hamiltonian. As it is discussed
in Ref.~\cite{av2}, it is the unitary representations
${\cal U}$ of the double cover $Spin(5)=Sp(2)$
of $SO(5)$ (alternatively the projective representations
of $SO(5)$) which define the time-dependent
Hamiltonian:
	\xbe
	\hat H'_0[R(t)]={\cal U}[g(t)]\: \hat H'_0[R(0)]\:
	 {\cal U}[g(t)]^\dagger\;.
	\label{time-dep}
	\xee
Here, $g:[0,T]\to Sp(2)$, is defined by $R(t)=:\pi[g(t)]
R(0)$, where $\pi :Sp(2)\to SO(5)$ is the canonical
two-to-one covering projection. The emergence of the
group $Sp(2)$ is an indication of the existence of a
quaternionic description of the system, \cite{av2}.

Let us next examine the situation for irreducible
representations $j$ of $SU(2)$. As I previously
described, $\hat J^2$ commutes with the Hamiltonian.
Hence the Hamiltonian is block-diagonal in the basis
with definite total angular momentum $j$.
For each $j$, the representation space ${\cal H}_j$ is
$2j+1$ dimensional. Therefore the restriction of the
Hamiltonian $\hat{H'}_0$ to  ${\cal H}_j$ and
${\cal U}[g(t)]$ are $(2j+1)\times(2j+1)$ matrices.

Let $\{ \phi^I_{j_n}\}$ be a complete set of orthonormal
eigenvectors of the initial Hamiltonian $\hat H'_0[R(0)]$,
where $I$ is a degeneracy label. Then the eigenvectors
of $\hat H'_0[R]$ are of the form:
	\xbe
	\Phi^I_{j_n}[R]={\cal U}[g]\:\phi^I_{j_n}\;,
	\label{phi=u-phi}
	\xee
and the non-Abelian connection one-form
(\ref{connection}) is given by
	\xbe
	{\cal A}_{j_n}^{IJ}=i\xbr \phi^I_{j_n}|\:
	{\cal U}^\dagger
	d\,{\cal U}\:|\phi^J_{j_n}\xkt\;,
	\label{be-co-bc}
	\xee
where $d$ stands for the exterior derivative operator
with respect to the parameters of the system.

For the integer $j$, bosonic systems, it is known that
the quadratic Hamiltonians of the form (\ref{h'0}), describe
time-reversal-invariant systems. In this case it can be
shown that the curvature two-form associated with the
Abelian Berry connection one-form (\ref{be-co})
vanishes identically \cite{av2}. The connection
one-form is exact (gauge potential is pure gauge)
and a nontrivial geometric phase can only be
topological, namely it may still exist provided
that the first homology group of the parameter space
is nontrivial.  For the problem under investigation
${\cal M}=S^4$, and the first homology group is
trivial. Hence, in general, Abelian geometric
phases are trivial. The same conclusion cannot
however be reached for the non-Abelian (matrix-valued)
geometric phases.

In the remainder of this section, I shall examine
the situation for some small values of $j$:
	\begin{itemize}
	\item[1)] $j=0$: The corresponding Hilbert subspace
is one-dimensional. Therefore, there is no nontrivial
geometric phases.
	\item[2)] $j=1/2$: In this case, $\hat J_a=\sigma_a/2$,
where $\sigma_a$ are Pauli matrices. Using the well-known
anticommutation (Clifford algebra) relations $\{\sigma_a,
\sigma_b\}=2 \delta_{ab}$, one can easily show that in this
case
	\xbe
	\hat H'=\frac{1}{2}\sum_{a}g^{aa}(t)\:\hat I\;,
	\label{s=1/2}
	\xee
where $\hat I$ is the identity matrix. Therefore, the
eigenvectors $\Phi_n$ are constant ($g_{ab}$-independent),
the connection one-form (\ref{be-co-bc}) vanishes
and the geometric phase angle is again zero.
	\item[3)] $j=1$: In this case the Hilbert subspace is
three-dimensional. The Abelian geometrical phases are
trivial. The nontrivial matrix-valued geometrical phases may be
present, provided that the Hamiltonian has a degenerate
eigenvalue.  Using the ordinary $j=1$ matrix representations
of $\hat J_a$, one can easily express $\hat H'_0$ as a
$3\times 3$ matrix. It can then be checked that in the generic
case the eigenvalues of $\hat H'_0$ are not 
degenerate.\footnote{This is true for all integer $j$,
\cite{av2}.} However, there are cases for which a
degenerate eigenvalue is present. A simple example is the
Taub metric, $(g_{ab})={\rm diag}(g_{11},g_{22},g_{22})$,
which is a particular example of the mixmaster metric, 
\cite{ra-sh}. For the general mixmaster metric, the
eigenvalue problem can be easily solved. The
eigenvalues of the total Hamiltonian $\hat H'$ of
Eq.~(\ref{tilde-h}) are
	\[
	\frac{1}{g_{11}}+\frac{1}{g_{22}}
	\;,~~~~~\frac{1}{g_{11}}+\frac{1}{g_{33}}
	\;,~~~~~\frac{1}{g_{22}}+\frac{1}{g_{33}}\;.\]
Therefore the degenerate case corresponds to the
coincidence of at least two of $g_{aa}$'s, i.e.,
a Taub metric. However, even in the general mixmaster
case, one can find a constant ($g_{aa}$-independent)
basis which diagonalizes the Hamiltonian. Hence
the non-Abelian connection one-form vanishes
and the geometric phase is again trivial. This is not
however the case for general metrics. In Appendix~B,
it is shown that without actually solving the general
eigenvalue problem for the general Hamiltonian, one can
find the conditions on the metric which render 
at least one of the eigenvalues of the Hamiltonian
degenerate.  Here, I summarize the results. Using the
well-known matrix representations of the angular
momentum operators $\hat J_a$ in the $j=1$
representation \cite{schiff} one can write the
Hamiltonian (\ref{tilde-h}) in the form:
	\xbe
	\hat H'=\left(\begin{array}{ccc}
	t+2z&\xi^*&\zeta^*\\
	\xi&2(t+z)&-\xi^*\\
	\zeta&-\xi&t+2z\end{array}\right)\;,
	\label{tilde-h'-0}
	\xee
where
	\xbea
	t&:=&\frac{1}{2}\:(g^{11}+g^{22})-g^{33}\;,~~~~~
	z\::=\:g^{33}\;,\xnn\\
	\xi&:=&\frac{1}{\sqrt{2}}\:(g^{13}+ig^{23})\;,~~~~~
	\zeta\::=\: \frac{1}{2}\:(g^{11}-g^{22})+ig^{12}\;.
	\xnn
	\xeea
Then it can be shown (App.~B), that the necessary
and sufficient conditions for $\hat H'$ to have a
degenerate eigenvalue are 
	\begin{itemize}
	\item[I.] for $\zeta=0$:  $\xi=0$, in which case,
$\hat H'$ as given by Eq.~(\ref{tilde-h'-0}) is already
diagonal. The degenerate and non-degenerate eigenvalues
are $t+2z$ and  $2(t+z)$, respectively. In terms of the
components of the metric, these conditions can be written
as: $g_{11}=g_{22}$ and $g_{ab}=0$ if $a\neq b$. This
is a Taub metric which as discussed above does not
lead to a nontrivial  geometric phase.
	\item[II.] for $\zeta\neq 0$:  
	\xbe
	\zeta={\cal Z}\,e^{2i\theta}\;,~~~~~ 
	t={\cal Z}-|\xi|^2/{\cal Z}\;, 
	\label{pa-eqs}
	\xee
where $\exp[i\theta]:=\xi/|\xi|$ and  ${\cal Z}\in\xR-\{ 0\}$.
In this case the degenerate and
non-degenerate eigenvalues are $2({\cal Z}+z)-|\xi|^2/
{\cal Z}$ and $2(z-|\xi|^2/{\cal Z})$, respectively.
\end{itemize}
For the latter case, an orthonormal set of eigenvectors
is given by:
	\xbe
	v_1=\frac{1}{\sqrt{2}}\,\left(\begin{array}{c}
	e^{-2i\theta}\\0\\1
	\end{array}\right)\,,~~~~
	v_2=\frac{1}{\sqrt{1+2{\cal X}^2}}\,
	\left(\begin{array}{c}
	{\cal X}e^{-i\theta}\\1\\-{\cal X}e^{i\theta}
	\end{array}\right)\,,~~~~
	v_3=\frac{1}{\sqrt{2(1+2{\cal X}^2)}}\,
	\left(\begin{array}{c}
	-e^{-2i\theta} \\2{\cal X}e^{-i\theta}\\1
	\end{array}\right)\,,
	\label{v's}
	\xee
where ${\cal X}:=|\xi|/(2{\cal Z})$. In view of the general
argument valid for all non-degenerate eigenvalues,
the geometric phase associated with $v_3$ is trivial.
This can be directly checked by substituting
$v_3$ in the formula (\ref{be-co})
for the Berry connection one-form.  This leads, after
some algebra, to the surprisingly simple result ${\cal A}_{33}
:=i\xbr v_3|dv_3\xkt=d\theta$. Therefore, ${\cal A}_{33}$
is exact as expected, and the corresponding geometric
phase angle vanishes. Similarly one can compute the
matrix elements ${\cal A}_{rs}:=i\xbr v_r|dv_s\xkt$,
$r,s=1,2$, of the non-Abelian connection one-form.
The result is:
	\xbea
	{\cal A}&=&\left(\begin{array}{cc}
	1&{\cal F}\\
	{\cal F}^*&0
	\end{array}\right)\:\omega\;,
	\label{be-co-v's}\\
	{\cal F}&:=&\frac{2{\cal X}\,e^{i\theta}}{
	\sqrt{2(1+2{\cal X}^2)}}\:=\:
	\frac{2\epsilon\xi}{\sqrt{2+\left|\frac{\xi}{\zeta}
	\right|^2}}\:=\:\frac{\epsilon(g^{13}+ig^{23})}{
	\sqrt{1+\frac{(g^{13})^2+(g^{23})^2}{ (g^{11}-
	g^{22})^2+(2g^{12})^2}}}\;,\xnn\\
	\omega&:=&d\theta\:=\:\frac{g^{13}dg^{23}-
	g^{23}dg^{13}}{(g^{13})^2+(g^{23})^2}\;,\xnn
	\xeea
where $\epsilon:={\cal Z}/|\zeta|=\pm$.
As seen from Eq.~(\ref{be-co-v's}), ${\cal A}$ is
a $u(2)$-valued one-form, which vanishes
if  $g^{23}/g^{13}$ is kept constant during the evolution
of the universe. 

It is also worth mentioning that the requirement of
the existence of degeneracy is equivalent to
restricting the parameters of the system to
a two-dimensional subset of $S^4$. Thus, the
corresponding spectral bundle \cite{si,p6} is a $U(2)$
vector bundle over a two-dimensional parameter space
$\tilde{\cal M}$. The manifold structure of $\tilde{\cal M}$
is determined by Eqs.~(\ref{pa-eqs}).  In terms of the
parameters $R^A$ of (\ref{R's}), these equations are
expressed by
	\xbe
	R^5=f_1\,R^4\;,~~~~~R^1=f_2\,R^4+
	\frac{f_3}{R^4}\;,
	\label{r5-r1}
	\xee
where
	\xbea
	f_1&:=&\frac{2R^2R^3}{(R^2)^2-(R^3)^2}\;,~~~~~
	f_2\::=\:\pm\frac{(R^2)^2+(R^3)^2}{\sqrt{3}\,[
	(R^2)^2-(R^3)^2]}\;,\xnn\\
	f_3&:=&\mp\frac{(R^2)^2-(R^3)^2}{2\sqrt{3}}\;,~~~~~
	f_4\::=\:(R^2)^2+(R^3)^2\;.
	\xnn
	\xeea
Here $f_4$ is also introduced for future use. In addition
to (\ref{r5-r1}), one also has the condition $(R^A)\in S^4$.
If $S^4$ is identified with the round sphere, this condition
takes the form $\sum_A (R^A)^2=1$. Substituting
(\ref{r5-r1}) in this equation, one finds
	\xbe
	(1+f_2^2+f_3^2)(R^4)^4-(1-f_4-2f_2f_3)(R^4)^2
	+f_3^2=0\;.
	\label{r4=f'a}
	\xee
Eq.~(\ref{r4=f'a}) may be easily solved for $R^4$. This
yields:
	\xbe
	R^4=\pm\,\frac{3[(R^2)^2-(R^3)^2]^2}{8[(R^2)^2+
	(R^3)^2]^2}\:\left[ 1-\frac{2}{3}[(R^2)^2+(R^3)^2]
	\pm\sqrt{1-\frac{4}{3}[(R^2)^2+(R^3)^2]}\:\right]\;.
	\label{r4=}
	\xee
Note that the parameters $R^A$ are related to the
components of the inverse of the three-metric through
Eqs.~(\ref{R's}). Thus the parameter space
$\tilde{\cal M}$ is really a submanifold of the
corresponding minisuperspace.
Fig.~1 shows a three-dimensional plot of $R^4$ as a
function of $R^2$ and $R^3$, i.e., a plot of the
parameter space $\tilde{\cal M}$ as embedded in $\xR^3$.
Note that $R^2=\pm R^3$ renders $f_1$
and $f_2$ singular. The corresponding points which are
depicted as the curves along which the figure becomes
non-differentiable must be handled with care.
The smooth part of $\tilde{\cal M}$ consists of eight connected
components, each of which is diffeomorphic to an open
disk (alternatively $\xR^2$). 
	\item[4)] $j=3/2$: This case has been studied in
Refs.~\cite{av-sa-se-si,av2} in detail. Therefore I suffice
to note that it involves nontrivial geometric phases.
	\end{itemize}

\section{Conclusion}
In this article I showed that the two-component
formalism could be consistently used to investigate
the geometric phases associated with charged
Klein-Gordon fields. This formalism provides a precise
definition of the adiabatic approximation and 
allows Berry's  derivation of the adiabatic geometrical
phase to be applied to the relativistic Klein-Gordon
fields.  In particular, I showed that the computation
of the adiabatic geometric phase did not involve
the explicit construction of an inner product on the
space of the initial conditions, or alternatively the space
of solutions of the Klein-Gordon equation. It only 
required the inner product  structure of  the Hilbert space
$L^2(\Sigma_t)$. 

In non-relativistic  quantum mechanics, the necessary
and sufficient condition for the validity of the adiabatic
approximation, $\psi\approx e^{i\alpha}|n\xkt$,
is $\xbr m|\dot n\xkt \approx 0$ for $m\neq n$,
\cite{messiyah,p16}, where $|n\xkt$ are instantaneous
eigenvectors of the Hamiltonian. If the Hamiltonian is
not self-adjoint then the eigenvectors may not be
orthogonal. In this case this condition is generalized to
$\xbr n|n\xkt\xbr m|\dot n\xkt-\xbr m| n\xkt\xbr n|\dot n
\xkt\approx 0$. A direct generalization of the
ansatz $\psi\approx e^{i\alpha}|n\xkt$ within the two-component
formulation of the Klein-Gordon equation leads
to an additional condition on the energy eigenvalues, namely
$\frac{d (q E_n)}{dt}\approx 0$. If this condition is satisfied
then the evolution is said to be ultra-adiabatic. If this
condition fails to be fulfilled but the adiabaticity condition
(\ref{approx-ort-condition}) is satisfied, then the evolution is
said to be adiabatic. The expressions for the geometric 
phase for  the ultra-adiabatic and adiabatic evolutions are
identical. The only difference is in the dynamical part of
the phase. 

I employed the general results of the two-component
formulation to study adiabatic geometric phases
induced by a rotating magnetic field and a rotating cosmic
string. The results were in complete agreement with those of
the previous investigations \cite{an-ma,co-pi},  but the 
analysis was considerably simpler.

I also investigated the geometric phases induced
by spatially  homogeneous cosmological backgrounds. In this
case the freedom in the choice of $q$ turned out to simplify
the analysis. I  showed that the adiabatic geometric
phase angles always vanished for Bianchi type I models, whereas
non-Abelian adiabatic phases could occur for the Bianchi type IX
models. Particularly, interesting was the relationship
between the corresponding  induced Hamiltonians in the
Bianchi type IX models and the quadrupole Hamiltonians of the
molecular and nuclear physics.  I also showed
that even for the integer spin representations nontrivial
geometric phases could exist. This should also be
of interest for the molecular physicists and chemists
who have apparently investigated only the fermionic
systems  (half-integer spin representations.)

In the context of general relativity where
the Poincar\'e invariance is replaced by the diffeomorphism
invariance, one can use the time-reparameterization symmetry
of the background gravitational field and the geometric phase
to absorb the magnitude $|q|$ of the decomposition parameter
$q$ into the definition of the lapse function $N=(-g^{00})^{-1/2}$.
In this way only a $U(1)$ subgroup of the corresponding
$GL(1,\xC)$ symmetry group survives. The additional requirement
of the self-adjointness of the Hamiltonian with respect to the
Klein-Gordon inner product, can then be used to choose
$q$ to be imaginary, i.e., to set $q=i$ or $q=-i$.  Note however that the
Klein-Gordon inner product does not play any distinguished
role in the computation of the geometric phase. Hence the
choice $q=i$ or $q=-i$ is not a necessary condition. It should also
be clear that the $GL(1,\xC)$ or $U(1)$ symmetry associated
with the freedom of choice of the decomposition has no
physical basis or consequences. It is merely a mathematical
feature of the two-component formalism which can occasionally
be used to simplify the calculations.

Finally, I wish to emphasize that the use of the two-component 
formulation in the study of the geometric phases associated with
scalar fields is more advantageous than the more conventional
approaches which are based on a decomposition of the
space of solutions into  positive and negative frequency
subspaces and the construction of a positive definite inner product,
e.g., those used in Refs.~\cite{an-ma,co-pi}.  This has two reasons.
Firstly, the conventional methods have apparently missed the fact
that one does not need to construct an inner product on the
space of Klein-Gordon fields to be able to calculate the adiabatic
geometric phase. Hence, a major part of these analyses is 
concerned with the construction of such an inner product. 
Secondly, these approaches can only be applied to the stationary
spacetimes where such an inner product can be constructed. 
The application of the two-component formulation for the
Bianchi models manifestly shows that this method can also be
employed even if the background spacetime is not
stationary.  One must however realize that similarly to the
conventional methods the present analysis is only valid within
the framework of  the adiabatic approximation. Although, the (approximate)
stationarity of  the background  metric (for  $A=0=V$) is a
sufficient condition for the validity of the adiabatic approximation,
it is not necessary. This can be easily seen by noting that
for example in the case of Bianchi  IX model, for spin $j=1/2$
states, one has $\dot\Phi_n=0$, so $\xbr \Phi_m|\dot\Phi_n\xkt=0$.
Therefore, although the spacetime is not stationary,
the adiabatic approximation yields the exact
solution of the field equation. This shows that in general for
arbitrary non-stationary spacetimes, there may exist adiabatically
evolving states to which the above analysis applies. This is in
contrast with the traditional methods of the computation of
the adiabatic geometric phases for scalar fields which 
involve construction of a Hilbert space structure on the space of
solutions of the Klein-Gordon equation.

\section*{Acknowledgements}
I would like to thank Bahman Darian for many fruitful
discussions and helping me with the computer graphics, 
Teoman Turgut for bringing to my attention one of the
references, and Kamran Saririan for mailing me copies
of a couple of the references.

\section*{Appendix~A}

In this Appendix I show that for the problem of a rotating
cosmic string, there exist localized energy eigenvectors.

Let $f=f(\rho,\varphi)$ be an arbitrary localized smooth
function, i.e., it has a compact support, and write
	\xbe
	f(\rho,\varphi)=\sum_m\int dk~\tilde f_m(k)\:
	e^{im\varphi}J_\nu(k\rho)\;,
	\label{f}
	\xee
where $\nu=m/\alpha$.
Using the orthogonality properties of the Bessel
functions and the `plane waves', Eq.~(\ref{f})  may be easily
inverted to yield $\tilde f_m(k)$ which in turn define a
localized field with definite energy eigenvalue $E_n$
according to:
	\[
	\Phi_n=e^{-4iE_nj\varphi}
	\sum_{m}\int dk~\tilde f_m(k)\: e^{im\varphi} 
	e^{i(\sqrt{E_n^2-\mu^2-k^2})z}J_\nu(k\rho) \;.
	\]
By construction $\Phi_n$ is localized in the $\rho$
and $\varphi$-directions. The localization in
$z$-direction is irrelevant to the arguments used
in section~5. Furthermore, note that there may still
be degenerate degrees of freedom left in this
construction.

\section*{Appendix~B}
In this Appendix I show how one can obtain the conditions
under which the Hamiltonian (\ref{tilde-h'-0}) has
degenerate eigenvalues without actually solving the
eigenvalue problem in the general case.

The analysis can be slightly simplified if one writes the
Hamiltonian (\ref{tilde-h'-0}) in the form:
	\xbea
	\hat H'&=&(t+2z)\hat I+\hat{\tilde H}\;,\xnn\\
	\hat{\tilde H}&:=&\left(\begin{array}{ccc}
	0&\xi^*&\zeta^*\\
	\xi&t&-\xi^*\\
	\zeta&-\xi&0\end{array}\right)\;,
	\label{tilde-h'}
	\xeea
where $\hat I$ is the $3\times 3$ identity matrix.  Clearly,
the eigenvalue problems for $\hat H'$ and $\hat{\tilde H}$
are equivalent. Computing the characteristic polynomial
for $\hat{\tilde H}$, i.e., $P(\lambda):=\det(\hat{\tilde H}-
\lambda \hat I)$, one finds:
	\xbe
	P(\lambda)=-\lambda^3+t\lambda^2+
	(|\zeta|^2+2|\xi|^2)\lambda-(t|\zeta|^2+\zeta\xi^{*2}
	+\zeta^*\xi^2)\;.
	\label{p}
	\xee
If one of the eigenvalues (roots of $P(\lambda)$) is
degenerate, then
	\xbe
	P(\lambda)=-(\lambda-l_1)(\lambda-
	l_2)^2\;.
	\label{p'}
	\xee
Comparing Eqs.~(\ref{p}) and (\ref{p'}), one finds
	\xbe
	t=l_1+2l_2\;,~~~l_2^2+2l_1l_2=-(|\zeta|^2+
	2|\xi|^2)\;,~~~l_1l_2^2=-(t|\zeta|^2+\zeta\xi^{*2}+
	\zeta^8\xi^2)\;.
	\label{1}
	\xee
Furthermore since $l_2$ is at least doubly degenerate,
the rows of the matrix:
	\xbe
	\hat{\tilde H}-l_2\: \hat I=
	\left(\begin{array}{ccc}
	-l_2&\xi^*&\zeta^*\\
	\xi&t-l_2&-\xi^*\\
	\zeta&-\xi&-l_2\end{array}\right)\;,
	\label{cof}
	\xee
must be mutually linearly dependent. In other words the
cofactors of all the matrix elements must vanish.
Enforcing this condition for the matrix elements
and using Eqs.~(\ref{1}), one finally finds that
either $\xi=\zeta=l_2=0$ and $l_1=t$, or 
	\xbea
	\zeta&=&{\cal Z}\,e^{2i\theta}\;,~~~~~
	 t \:=\:{\cal Z}-\frac{|\xi|^2}{{\cal Z}}\;, \xnn\\
	l_2&=&{\cal Z}\;,~~~~~~l_1\:=\:-({\cal Z}+
	\frac{|\xi|^2}{{\cal Z}})\;,\xnn
	\xeea
where $\exp[i\theta]:=\xi/|\xi|$ and  ${\cal Z}\in\xR-\{ 0\}$.

\newpage
.
	\begin{figure}
	\caption{\small This is a plot of $R^4=R^4(R^2,R^3)$.
	The horizontal plane is the $R^2$-$R^3$-plane
	and the vertical axis is the $R^4$-axis. The
	parameter space $\tilde{\cal M}$ is obtained
	by removing the intersection of this figure with
	the planes defined by: $R^4=0$, $R^2=R^3$ and
	$R^2=-R^3$. The intersection involves the curves
	along which the figure becomes
	non-differentiable.}
	\label{fig}
	\end{figure} 
\newpage

\thispagestyle{empty}
\begin{figure}
\epsffile{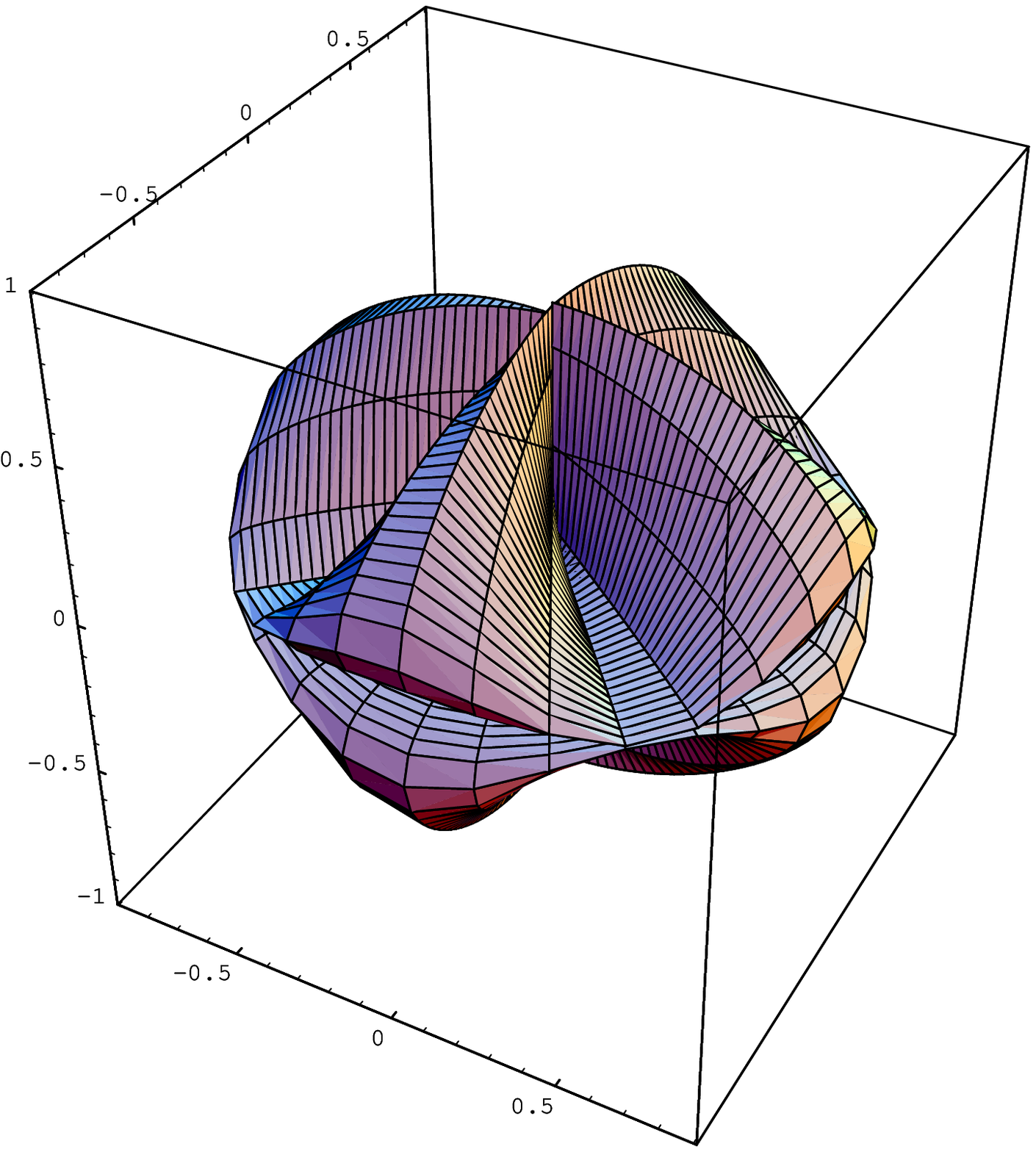}
\end{figure}

\end{document}